\documentclass[aps,12pt,showpacs,preprint,groupedaddress,floatfix,nofootinbib]{revtex4}

\usepackage{}
\usepackage{graphicx}
\usepackage{dcolumn}
\usepackage{bm}
\usepackage{amssymb}
\usepackage{amsmath}
\usepackage{supertabular}
\usepackage{xcolor}
\usepackage{mathrsfs}
\usepackage{fontenc}

\begin{document}

\title{Microscopic core-quasiparticle coupling model for spectroscopy of odd-mass nuclei}
\author{S. Quan$^{1}$}
\author{W. P. Liu$^{1}$}
\author{Z. P. Li$^{1,2}$}\email{zpliphy@swu.edu.cn}
\author{M. S. Smith$^{2}$}\email{smithms@ornl.gov}

\affiliation{$^{1}$School of Physical Science and Technology, Southwest University, Chongqing 400715, China}
\affiliation{$^{2}$Physics Division, Oak Ridge National Laboratory, Oak Ridge, Tennessee, 37831-6354, USA}

\bigskip
\date{\today}

\begin{abstract}
\begin{description}
\item[Background] Predictions of the spectroscopic properties of low-lying states are critical for nuclear structure studies, but are problematic for nuclei with an odd nucleon due to the interplay of the unpaired single particle with nuclear collective degrees of freedom.

\item[Purpose] To predict the spectroscopic properties of odd-mass medium-heavy and heavy nuclei with a model that treats single-particle and collective degrees of freedom within the same microscopic framework.

\item[Method] A microscopic core-quasiparticle coupling (CQC) model based on the covariant density functional theory is developed that contains the collective excitations of even-mass cores and spherical single-particle states of the odd nucleon as calculated from a quadrupole collective Hamiltonian combined with a constrained triaxial relativistic Hartree-Bogoliubov model.

\item[Results] Predictions of the new model for excitation energies, kinematic and dynamic moments of inertia, and transition rates are shown to be in good agreement with results of low-lying spectroscopy measurements of the axially deformed odd-proton nucleus $^{159}$Tb and the odd-neutron nucleus $^{157}$Gd.

\item[Conclusions] A microscopic CQC model based on covariant density functional theory is developed for odd-mass nuclei and shown to give predictions that agree with measurements of two medium-heavy nuclei. Future studies with additional nuclei are planned.

\end{description}
\end{abstract}


\maketitle

\section{\label{secI}Introduction}

The nuclear spectroscopic properties of low-lying states are important physical quantities that
reveal rich structure information of atomic nuclei, including shape phase transitions,
evolution of the shell structure, isomeric states, shape coexistence, and more \cite{Bohr75,Ring80,Cejnar10,Heyde11}.
Since nuclei are finite-size, strong correlated quantal many-body systems,
their complex spectra exhibit a large variety of excitation modes that relate to either
collective or single-particle degrees of freedom, or the coupling between them \cite{Bohr75,Ring80}.

Global, microscopic descriptions of complex nuclear spectra require modeling the in-medium nucleon-nucleon interaction.
Here we focus on methods based on an energy density functional (EDF),
which have been successfully used over the whole nuclide chart \cite{BHR.03,VALR.05,Meng06,Stone07,Meng16,Meng06b}.
In general, frameworks based on static nuclear mean field approximations
can only describe ground-state properties such as binding energies and charge radii.
Calculating excitation spectra and electromagnetic transition probabilities requires including correlations
beyond the static mean field through the restoration of broken
symmetries and configuration mixing of symmetry-breaking
product states. The most effective approach to configuration
mixing calculations is the generator coordinate method (GCM) \cite{Ring80},
with multipole moments used as collective coordinates
to generate symmetry-breaking product wave functions.
As the Gaussian overlap approximation of GCM,
quadrupole collective Hamiltonians with parameters determined by self-consistent
mean-field calculations are numerically simpler,
and have achieved great success in the description of low-lying states
in a wide range of nuclei, from $A\sim40$ to superheavy regions including
both stable and unstable nuclei \cite{Libert99,Pro.99,Nik.09,Li.09,Li.10,Li.11,Li12,Lu15,Del10,Nik11,Quan17}.
The validity of this approximate method was recently demonstrated by a comparison with a full GCM
calculation for a shape coexisting nucleus $^{76}$Kr based on a covariant EDF \cite{Yao14}.

Most studies using the GCM or quadrupole collective Hamiltonians based on EDFs are limited to even-even nuclei.
Calculations for odd-mass nuclei are much more complicated  due to the interplay
between the unpaired single-particle and collective degrees of freedom.
Recently, two EDF-based approaches have been reported for odd-mass nuclei.
One is a consistent extension of GCM,
where the generator coordinate space is built on blocked one-quasiparticle
Hartree-Fock-Bogoliubov (HFB) states.
In Ref. \cite{Bally14}, a fully GCM calculation based on angular-momentum and particle-number projected triaxially
deformed HFB states using the Skyrme SLyMR0 parametrization was performed for the low-lying spectrum of
$^{25}$Mg. Refs. \cite{Borrajo16,Borrajo17} presented an approach for the calculation of odd nuclei with exact self-consistent blocking and particle number and angular-momentum projection with the finite-range density-dependent Gogny force as applied to the study of Mg isotopic chain.  The other approach is the beyond-mean-field boson-fermion model based on the framework of nuclear energy density functional theory \cite{Nomura16}. This method uniquely determines the parameters of the Hamiltonian of the boson core, while the strength of the particle-core coupling is specifically adjusted to selected data for a particular nucleus. The approach is illustrated
in a systematic study of low-energy excitation spectra and transition rates of axially deformed odd-mass Eu isotopes.

In this work, our covariant EDF-based quadrupole collective Hamiltonian will be extended to describe
the spectroscopy of odd-mass nuclei via the core-quasiparticle coupling (CQC) scheme.
The CQC scheme has been extensively used with phenomenological inputs, {\it e.g.},
a rotor or Bohr Hamiltonian for the core and a single particle in a phenomenological spherical potential
\cite{Donau77,Donau79,Prot94,Prot96,Prot97,Klein00,Klein04,Droste04}, and microscopic inputs calculated from Hartree-Fock plus BCS \cite{Meyer79,Medjadi86,Libert82}.
Here, we will construct a microscopic CQC model, where the  collective degrees of freedom of the core
and single-particle will be both treated within the same covariant EDF.
The framework of the present model is similar as that of the previous work with microscopic inputs \cite{Meyer79,Medjadi86,Libert82}, while in our model a fully microscopic quadrupole collective Hamiltonian is used to describe the core. Also,  the inclusion of neighboring even cores enables the model to take into account the shape polarization effect that is critical for transitional nuclei.
We utilize the advantages of the EDF-based quadrupole collective Hamiltonian for even-even nuclei,
namely a clear physical picture constructed from the concept of nuclear shapes,
a global model that can be used for both stable and unstable nuclei, and reasonable computational speed for heavy nuclei.

In Section \ref{secII}, we describe the method to construct the CQC Hamiltonian and  the calculations of
the microscopic inputs. In Section~\ref {secIII}, the model is tested in a series of illustrative calculations
for the spectroscopic properties of the axially deformed odd-proton nucleus $^{159}$Tb and the odd-neutron nucleus $^{157}$Gd.
Section \ref{secIV} contains a summary of results and an outlook for future studies.

\section{\label{secII} Theoretical framework}
\subsection{Core-quasiparticle coupling model}

In the core-quasiparticle coupling scheme, the odd-mass nucleus with mass number $A$ is considered to be composed of both
a particle coupled to the lighter even neighbor $A-1$ and a hole coupled to the heavier even neighbor $A+1$.
The ansatz of the wave function for odd-mass nucleus can therefore be written as
\begin{eqnarray}
\label{eq:wf}
|\alpha JM_J\rangle^A &=& \sum_{\mu j,\nu R} \Big\{ U_{\alpha J}(\mu j,\nu R)\big[a^\dag_{\mu jm_j}|\nu RM_R\rangle\big]^{A-1}_{JM_J} \nonumber \\
                         &  & \ \ \ \ +V_{\alpha J}(\mu j,\nu R)\big[a_{\mu jm_j}|\nu RM_R\rangle\big]^{A+1}_{JM_J}\Big\}
\end{eqnarray}
where $\alpha$ denotes all quantum numbers beside the total angular momentum $J$ and its projection $M_J$ for the odd-$A$ nucleus.
$\mu$ and $\nu$ play the same roles as $\alpha$ but for the single-particle states and collective states of the core, respectively.
The linear coefficients $U_{\alpha J}(\mu j,\nu R)$ and $V_{\alpha J}(\mu j,\nu R)$ represent the probability amplitudes
for the particle-like and hole-like states, respectively,
which are formed by vector-coupling of a spherical particle state $|\mu jm_j\rangle^{A-1}$ to a collective state $|\nu RM_R\rangle^{A-1}$ of the core $A-1$ and a spherical hole state $|\mu j\bar{m}_j\rangle^{A+1}$ coupled to the corresponding collective state $|\nu RM_R\rangle^{A+1}$ of the core $A+1$, respectively.

The Hamiltonian for CQC model can be written in a general form\footnote{In the present work, the expression of CQC Hamiltonian is a little bit different from that in the Eqs. (34, 35) of Ref. \cite{Klein04} because the signs of our definitions for pairing field $\Delta$ and eigenvalue $E_{\alpha J}$ are opposite to those in Ref. \cite{Klein04}.} \cite{Klein04}
\begin{align}
H &=H_{\rm qp}+H_{\rm c}  \nonumber\\
   &=\left(
\begin{array}{cc}
(\varepsilon^{A-1}-\lambda)+\Gamma^{A-1} & \Delta^{A+1} \\
\Delta^{\dag A-1} & -(\varepsilon^{A+1}-\lambda)-\Gamma^{A+1}
\end{array}\right)
+\left(
\begin{array}{cc}
E^{A-1} & 0 \\
0 & E^{A+1}
\end{array}\right).
\label{eq:Ham}
\end{align}
The matrices $(\varepsilon^{A\pm1}-\lambda)$ and $(E^{A\pm1})$ are diagonal with respect to the basis states in the decomposition (\ref{eq:wf})
\begin{align}
(\varepsilon^{A\pm1}-\lambda) &= (\varepsilon_{\mu j}^{A\pm1}-\lambda)\delta_{\mu j,\mu^\prime j^\prime}\delta_{\nu R,\nu^\prime R^\prime} \\
(E^{A\pm1}) &= E_{\nu R}^{A\pm1}\delta_{\mu j,\mu^\prime j^\prime}\delta_{\nu R,\nu^\prime R^\prime}
\end{align}
with the single-particle energies $\varepsilon_{\mu j}^{A\pm1}$, Fermi surface $\lambda$, and collective excitation energies $E_{\nu R}^{A\pm1}$.
$\Gamma$ and $\Delta$ are the mean field and pairing field related to the long-range particle-hole interaction
and short-range particle-particle interaction between the odd nucleon and core, respectively.
In the present version of the model, the dominate quadrupole-quadrupole interaction and monopole pairing force
are used to determine the fields $\Gamma$ and $\Delta$, respectively \cite{Klein04}
\begin{align}
\label{eq:Gamm}
\left(\Gamma^{A\pm1}\right) &=-\chi (-1)^{j+R+J}\left\{\begin{array}{ccc} j & 2 & j^\prime \\ R^\prime & J & R \end{array}\right\}
                              \langle\mu j\|\hat Q_2\|\mu^\prime j^\prime\rangle^{A\pm1} \langle\nu R\|\hat Q_2\|\nu^\prime R^\prime\rangle^{A\pm1} \\
\left(\Delta^{A+1}\right) &=\left(\Delta^{A-1}\right) =\langle\nu R;A-1|\hat\Delta|\nu^\prime R^\prime;A+1\rangle \delta_{\mu j,\mu^\prime j^\prime}
            \approx \frac{1}{2}(\Delta^{A-1}_{\nu R}+\Delta^{A+1}_{\nu R})
            \delta_{\mu j,\mu^\prime j^\prime}\delta_{\nu R,\nu^\prime R^\prime}\equiv\left(\Delta\right)
\end{align}
where $\langle\mu j\|\hat Q_2\|\mu^\prime j^\prime\rangle^{A\pm1}$ and $\langle\nu R\|\hat Q_2\|\nu^\prime R^\prime\rangle^{A\pm1}$ are the reduced quadrupole matrix elements of the spherical hole (particle) and cores, respectively, while $\chi$ is the coupling strength of the quadrupole field. $\Delta_{\nu R}^{A\pm1}$ denotes the average pairing gaps of the collective states $|\nu RM_R\rangle^{A\pm1}$.

In the present work, the Fermi surface $\lambda$ and coupling strength $\chi$ are left as free parameters that are fit to data
separately for positive- and negative-parity states.
Finally, we obtain the excitation energies $E_{\alpha J}$ and linear coefficients $U_{\alpha J}(\mu j,\nu R)$, $V_{\alpha J}(\mu j,\nu R)$
in the wave functions of the odd-$A$ nucleus by solving the eigen equation
\begin{equation}
\label{eq:hamo}
\left(
\begin{array}{cc}
(\varepsilon^{A-1}-\lambda)+\Gamma^{A-1}+E^{A-1} & \Delta \\
\Delta & -(\varepsilon^{A+1}-\lambda)-\Gamma^{A+1}+E^{A+1}
\end{array}\right)
\left(
\begin{array}{c}
U \\ V
\end{array}\right)
=E_{\alpha J}
\left(
\begin{array}{c}
U \\ V
\end{array}\right)
\end{equation}
following the method introduced in Ref.~\cite{Klein04}.

In core-quasiparticle coupling scheme, the electromagnetic multipole operator is composed of
two parts contributed from the single particle and even-mass core
\begin{equation}
\hat{M}_{\lambda\mu}=\hat{M}^{\text{s.p.}}_{\lambda\mu}+\hat{M}^{\text{c}}_{\lambda\mu}.
\end{equation}
The corresponding reduced transition matrix element between states $|\alpha_1J_1\rangle$
and $|\alpha_2J_2\rangle$ reads
\begin{align}
 \langle \alpha_1J_1\|\hat{M}_{\lambda}\|\alpha_2J_2\rangle
 &=\sqrt{(2J_1+1)(2J_2+1)}\sum_{\nu R,\mu_1j_1\mu_2j_2}(-1)^{j_1+J_2+R+\lambda}
 \left\{\begin{array}{ccc}
         J_1 & \lambda & J_2 \\
         j_2 & R       & j_1
 \end{array}\right\}
 \Big[{M}^{\text{s.p.}}_U+{M}_V^{\text{s.p.}}\Big]\nonumber \\
&+\sqrt{(2J_1+1)(2J_2+1)}\sum_{\mu j,\nu_1R_1\nu_2R_2}(-1)^{R_1+J_2+j+\lambda}
 \left\{\begin{array}{ccc}
         J_1 & \lambda & J_2 \\
         R_2 & j       & R_1
 \end{array}\right\}
\Big[{M}^{\text{c}}_U+{M}_V^{\text{c}}\Big]
\end{align}
with
\begin{align}
\label{eq:MUsp}
M_U^{\text{s.p.}}&=U_{\alpha_1J_1}(\mu_1j_1,\nu R)U_{\alpha_2J_2}(\mu_2 j_2,\nu R)
                                \langle \mu_1j_1\|\hat{M}^{\text{s.p.}}_{\lambda}\|\mu_2j_2\rangle^{A-1} \\
M_V^{\text{s.p.}}&=V_{\alpha_1J_1}(\mu_1j_1,\nu R)V_{\alpha_2J_2}(\mu_2 j_2,\nu R)
                                \langle \mu_1j_1\|\hat{M}^{\text{s.p.}}_{\lambda}\|\mu_2j_2\rangle^{A+1} \\
M_U^{\text{c}}   &=U_{\alpha_1J_1}(\mu j,\nu_1R_1)U_{\alpha_2J_2}(\mu j,\nu_2R_2)
                               \langle \nu_1R_1\|\hat{M}^{\text{c}}_{\lambda}\|\nu_2R_2\rangle^{A-1} \\
\label{eq:MVc}
M_V^{\text{c}}   &=V_{\alpha_1J_1}(\mu j,\nu_1R_1)V_{\alpha_2J_2}(\mu j,\nu_2R_2)
                               \langle \nu_1R_1\|\hat{M}^{\text{c}}_{\lambda}\|\nu_2R_2\rangle^{A+1}.
\end{align}
The reduced matrix elements for electric quadrupole transitions and magnetic dipole transitions will be presented in Sec. \ref{SecIIB} in detail.

\subsection{\label{SecIIB} Microscopic inputs based on covariant EDF}

The full dynamics of CQC Hamiltonian Eq. (\ref{eq:Ham}) is determined by the energies $\varepsilon_{\mu j}^{A\pm1}$ and $E^{A\pm1}_{\nu R}$, quadrupole matrix elements   $\langle\mu j\|\hat Q_2\|\mu^\prime j^\prime\rangle^{A\pm1}$ and $\langle\nu R\|\hat Q_2\|\nu^\prime R^\prime\rangle^{A\pm1}$, and pairing gaps $\Delta_{\nu R}^{A\pm1}$ corresponding to the spherical hole (particle) states of the odd nucleon and collective excitation states of the cores. In the following, the superscript $A\pm1$  will be omitted for convenience.  In this part, we will calculate all the inputs for CQC model from a triaxial relativistic Hartree-Bogoliubov (RHB) model combined with a quadrupole collective Hamiltonian \cite{Nik14,Li.10,Li.11}.
The RHB model provides a unified description of particle-hole $(ph)$ and particle-particle
$(pp)$ correlations on a mean-field level by combining two average potentials: the self-consistent mean field that
encloses long range \textit{ph} correlations, and a pairing field $\hat{\Delta}$ which sums up
\textit{pp}-correlations.  In the present analysis, the mean-field potential is determined by the relativistic density functional PC-PK1 \cite{Zhao10} in the $ph$ channel, and a separable pairing
force~\cite{TMR.09a,Niksic10} is used in the $pp$ channel.

In the first step of the construction of CQC Hamiltonian (\ref{eq:Ham}),
a constrained RHB calculation for a spherical configuration of the even-mass cores are made
to obtain the single-particle energies $\varepsilon_{\mu j}$,
wave functions $|\mu jm_j\rangle$, and quadrupole matrix elements
\begin{align}
\langle\mu j\|\hat Q_2\|\mu^\prime j^\prime\rangle &=\langle\mu j\|r^2Y_2\|\mu^\prime j^\prime\rangle\nonumber\\
      &=(-1)^{j+j^\prime+1}\sqrt{\frac{5(2j^\prime+1)}{4\pi}}C^{j\frac{1}{2}}_{j^\prime\frac{1}{2} 20}\langle\mu j|r^2|\mu^\prime j^\prime\rangle
\label{eq:spq}
\end{align}
where $C^{j\frac{1}{2}}_{j^\prime\frac{1}{2} 20}$ are the Clebsch-Gordan coefficients.

Secondly, a constrained RHB calculation for the entire energy surface as functions of the quadrupole deformation $\beta$ and $\gamma$ is performed to provide the microscopic inputs, {\it i.e.}, the moments of inertia $\mathcal{I}_k\ (k=1, 2, 3)$, collective masses $B_{\beta\beta}$, $B_{\beta\gamma}$, $B_{\gamma\gamma}$, and the potential $V_{\textnormal{coll}}$, for the quadrupole collective Hamiltonian describing the collective vibration, rotation, and the coupling between them of the even-mass core \cite{Li.10,Li.11}
\begin{align}
\hat{H}_{\rm coll} =&-\frac{\hbar^2}{2\sqrt{wr}}
   \left\{\frac{1}{\beta^4}
   \left[\frac{\partial}{\partial\beta}\sqrt{\frac{r}{w}}\beta^4
   B_{\gamma\gamma} \frac{\partial}{\partial\beta}
   - \frac{\partial}{\partial\beta}\sqrt{\frac{r}{w}}\beta^3
   B_{\beta\gamma}\frac{\partial}{\partial\gamma}
   \right]\right.
   \nonumber \\
   &+\frac{1}{\beta\sin{3\gamma}}\left.\left[
   -\frac{\partial}{\partial\gamma} \sqrt{\frac{r}{w}}\sin{3\gamma}
      B_{\beta \gamma}\frac{\partial}{\partial\beta}
    +\frac{1}{\beta}\frac{\partial}{\partial\gamma} \sqrt{\frac{r}{w}}\sin{3\gamma}
      B_{\beta \beta}\frac{\partial}{\partial\gamma}
   \right]\right\} \nonumber \\
   &+\frac{1}{2}\sum_{k=1}^3{\frac{\hat{J}^2_k}{\mathcal{I}_k}}+V_{\rm coll}\; .
\end{align}
$\hat{J}_k$ denotes the components of the angular momentum in
the body-fixed frame of a nucleus, and
the moments of inertia $\mathcal{I}_k$ depend on the quadrupole
deformation variables $\beta$ and $\gamma$:
\begin{equation}
\mathcal{I}_k = 4B_k\beta^2\sin^2(\gamma-2k\pi/3) \;.
\end{equation}
Two additional quantities that appear in the expression for the vibrational energy:
$r=B_1B_2B_3$ and $w=B_{\beta\beta}B_{\gamma\gamma}-B_{\beta\gamma}^2 $,
determine the volume element in the collective space.

The diagonalization of this Hamiltonian yields the excitation energies $E_{\nu R}$ and collective wave functions
\begin{equation}
|\nu RM_R\rangle=\sum\limits_K\psi^R_{\nu K}(\beta, \gamma)\Phi^R_{M_RK}(\Omega)
\end{equation}
where  $K$ is the projection of angular momentum $R$ on the third axis in the body-fixed frame
and $\Phi^R_{M_RK}(\Omega)$ is a linear combination of Wigner $D$ functions as functions of Euler angles $\Omega$.
Then we can calculate the reduced matrix elements
\begin{align}
\langle\nu R\|\hat Q_2\|\nu^\prime R^\prime\rangle&=\sqrt{2R+1}\sum\limits_K\int\beta^4|\sin3\gamma|d\beta d\gamma
\Big[C^{RK}_{R^\prime K20}\psi^{R}_{\nu K}\psi^{R^\prime}_{\nu^\prime K}q_{20}(\beta,\gamma) \nonumber\\
\label{eq:cq}
& \ \ \ +\sqrt{\frac{1+\delta_{K0}}{2}}(C^{RK+2}_{R^\prime K22}\psi^{R}_{\nu K}\psi^{R^\prime}_{\nu^\prime K+2}
         +C^{RK}_{R^\prime K+2 2-2}\psi^{R}_{\nu K}\psi^{R^\prime}_{\nu^\prime K-2})q_{22}(\beta,\gamma)\Big]\\
\Delta_{\nu R} &= \sum\limits_K\int\beta^4|\sin3\gamma|d\beta d\gamma |\psi^{R}_{\nu K}|^2\Delta(\beta,\gamma)
\end{align}
where $q_{20}(\beta,\gamma)$ ($q_{22}(\beta,\gamma)$) and $\Delta(\beta,\gamma)$ are the mass quadrupole moments
and pairing gaps calculated from the Slater determinant of the RHB model for each deformation value $(\beta,\ \gamma)$.

For the electric quadrupole transitions, the reduced matrix elements in Eqs. (\ref{eq:MUsp}-\ref{eq:MVc}) are
$\langle\mu_1 j_1\|\hat Q^p_2\|\mu_2 j_2\rangle$ and $\langle\nu_1 R_1\|\hat Q^p_2\|\nu_2 R_2\rangle$,
which have the same expressions as the quadrupole matrix elements in Eqs.~(\ref{eq:spq}, \ref{eq:cq}) in the special case of protons. For the magnetic dipole transitions, the reduced matrix elements for the single particle are calculated using a nonrelativistic approximation
\begin{align}
\langle \mu_1j_1\|\hat{M}^{\text{s.p.}}_{1}\|\mu_2j_2\rangle
&=\langle\mu_1 j_1\|(g_s\mathbf{s}+g_l\mathbf{l})\cdot(\nabla rY_1)\|\mu_2 j_2\rangle \nonumber\\
      &=(-1)^{j_1+j_2+1}\sqrt{\frac{3(2j_2+1)}{4\pi}}C^{j_1\frac{1}{2}}_{j_2\frac{1}{2} 10}\langle\mu_1 j_1|\mu_2 j_2\rangle
      (1-k)\left[\frac{1}{2}g_s-g_l(1+\frac{k}{2})\right]
\label{eq:M1sp}
\end{align}
with $k=(j_1+1/2)(-1)^{j_1+l_1+1/2}+(j_2+1/2)(-1)^{j_2+l_2+1/2}$.
Here, $g_s$ and $g_l$ are the $g$-factors for the spin and orbital parts of the single particle, respectively.
For the core,
\begin{equation}
\label{eq:M1c}
\langle \nu_1R_1\|\hat{M}^{\text{c}}_{1}\|\nu_2R_2\rangle
=\langle \nu_1R_1\|\hat{M}^{\text{c}}_{1}\|\nu_1R_1\rangle\delta_{\nu_1R_1,\nu_2R_2}
=\frac{\sqrt{2R_1+1}}{C^{R_1R_1}_{R_1R_110}}g_cR_1\delta_{\nu_1R_1,\nu_2R_2}
\end{equation}
where $g_c$ is the $g$-factor for the core.

\section{\label{secIII}Illustrative calculations for well-deformed odd-mass nuclei}

\begin{figure}[htb]
\includegraphics[scale=0.6]{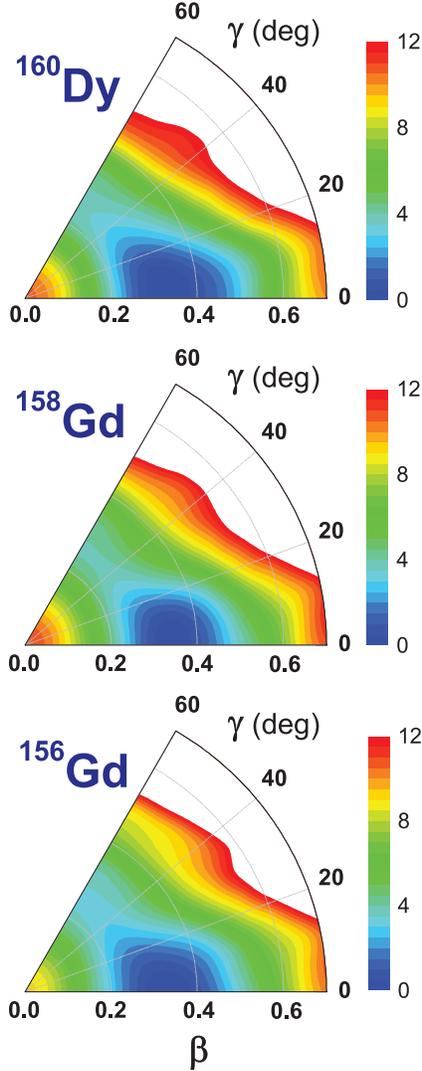}
\caption{\label{fig:PES}(Color online) Triaxial energy surfaces of
the core nuclei $^{160}$Dy, $^{158}$Gd, and $^{156}$Gd in the $\beta-\gamma$ plane
($0\le \gamma \le 60^0$) as calculated from a constrained triaxial RHB model.
For each nucleus, energies are normalized with respect to the
binding energy of the global minimum. The contours join points on
the surface with the same energy (in MeV).}
\end{figure}

\begin{figure}[htb]
\includegraphics[scale=0.4]{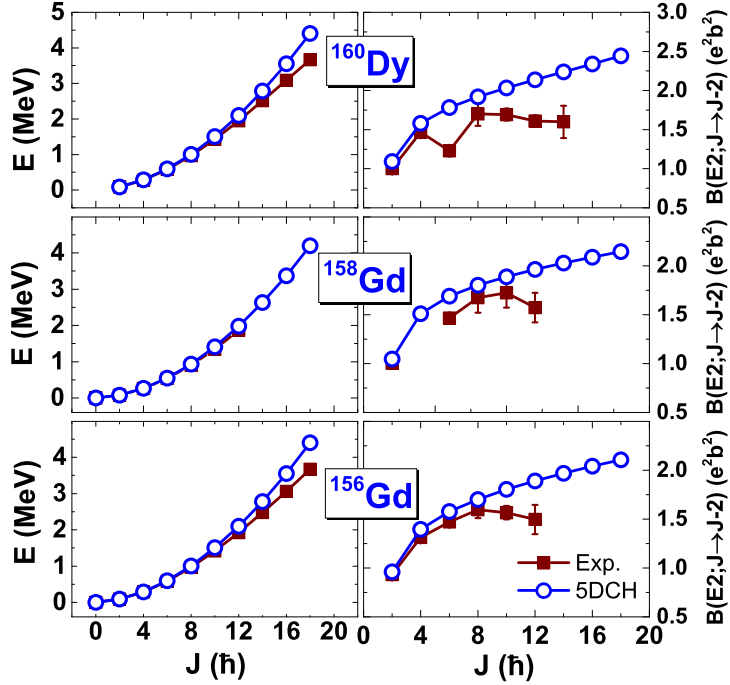}
\caption{\label{fig:Ecore}(Color online) The excitation energies (left panels) and reduced electric quadrupole transitions $B(E2)$ (right panels)
of the ground state bands in the core nuclei $^{160}$Dy, $^{158}$Gd, and $^{156}$Gd.
The theoretical results are obtained from the quadrupole collective Hamiltonian with collective parameters determined by
the constrained triaxial RHB model using the PC-PK1 density functional.
The experimental data are taken from Refs.~\cite{Reich12a,Helmer04,Lee85}.}
\end{figure}

As an illustrative application of the microscopic CQC model, we consider the case of a single nucleon coupled to
an axially symmetric rotor: the spectroscopy of the odd-proton nucleus $^{159}$Tb and the odd-neutron nucleus $^{157}$Gd.
We choose these because there is extensive data on their electromagnetic transition rates. The corresponding even-core nuclei
$^{160}$Dy, $^{158}$Gd, and $^{156}$Gd present excellent examples of axially-deformed rotors,
as shown by the potential energy surfaces in Fig.~\ref{fig:PES} as calculated from the constrained triaxial RHB model.
The axial deformation parameter $\beta$ is $\sim0.35$ for these three nuclei.
Solving the quadrupole collective Hamiltonian based on their potential energy surfaces yields the collective excitation states
of the even-core nuclei. Figure \ref{fig:Ecore} displays the resulting excitation energies
and intraband $B(E2)$ transitions of the ground state bands.
The theoretical results are in very good agreement with the experimental data, especially for low-lying states.


\begin{figure*}[htb]
\includegraphics[scale=0.5]{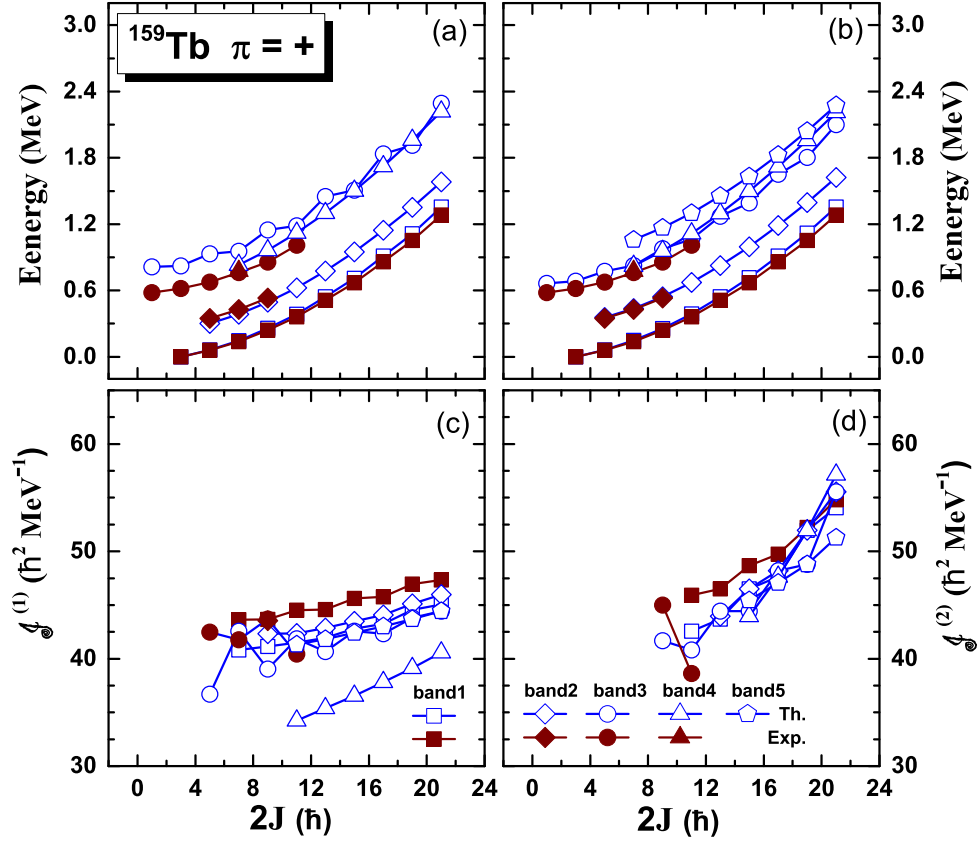}
\caption{(Color online) The calculated low-energy positive-parity bands (panels a, b), kinematic moments of inertia (panel c), and dynamic moments of inertia (panel d) of the odd-proton nucleus $^{159}$Tb, plotted in comparison with experimental data \cite{NNDC}. The theoretical results are calculated from CQC model with only the ground state band of the core (panel a) and with both the ground state and $\gamma$ bands of the core (panels b, c, d).  The Fermi surface and coupling strength ($\lambda$, $\chi$) in the CQC model are chosen as  (-8.50 MeV, 9.40 MeV/$b^2$) in the calculations.}
\label{fig:spec-Tb159-po}
\end{figure*}

\begin{figure*}[htb]
\includegraphics[scale=0.5]{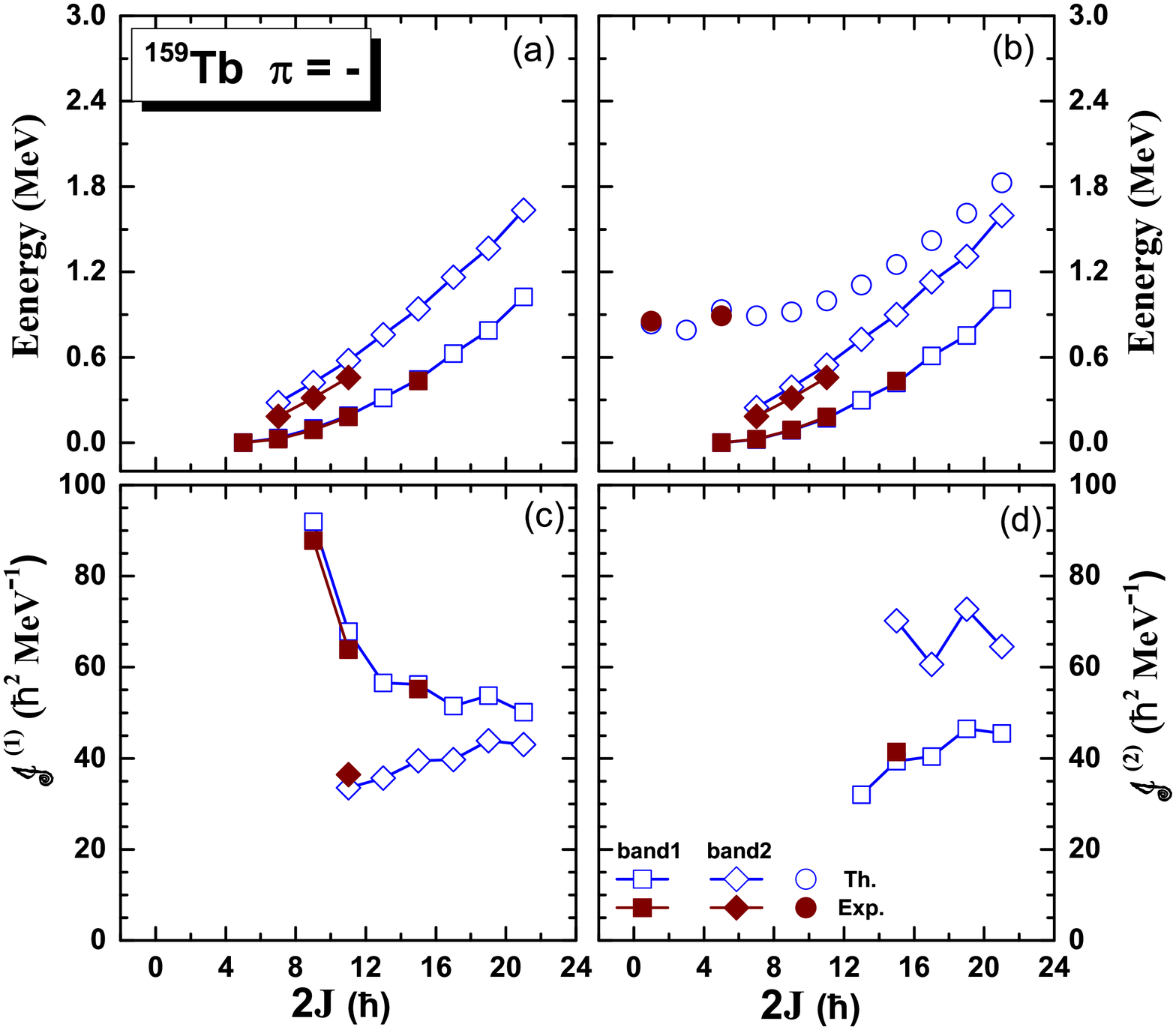}
\caption{(Color online) Same as Fig. \ref{fig:spec-Tb159-po} but for the negative-parity bands of $^{159}$Tb. The excitation energies in the upper panels are shown relative to the lowest state. The Fermi surface and coupling strength ($\lambda$, $\chi$) in the CQC model are chosen as  (-7.00 MeV, 12.80 MeV/$b^2$). }
\label{fig:spec-Tb159-Ne}
\end{figure*}


\begin{figure*}[htb]
\includegraphics[scale=0.5]{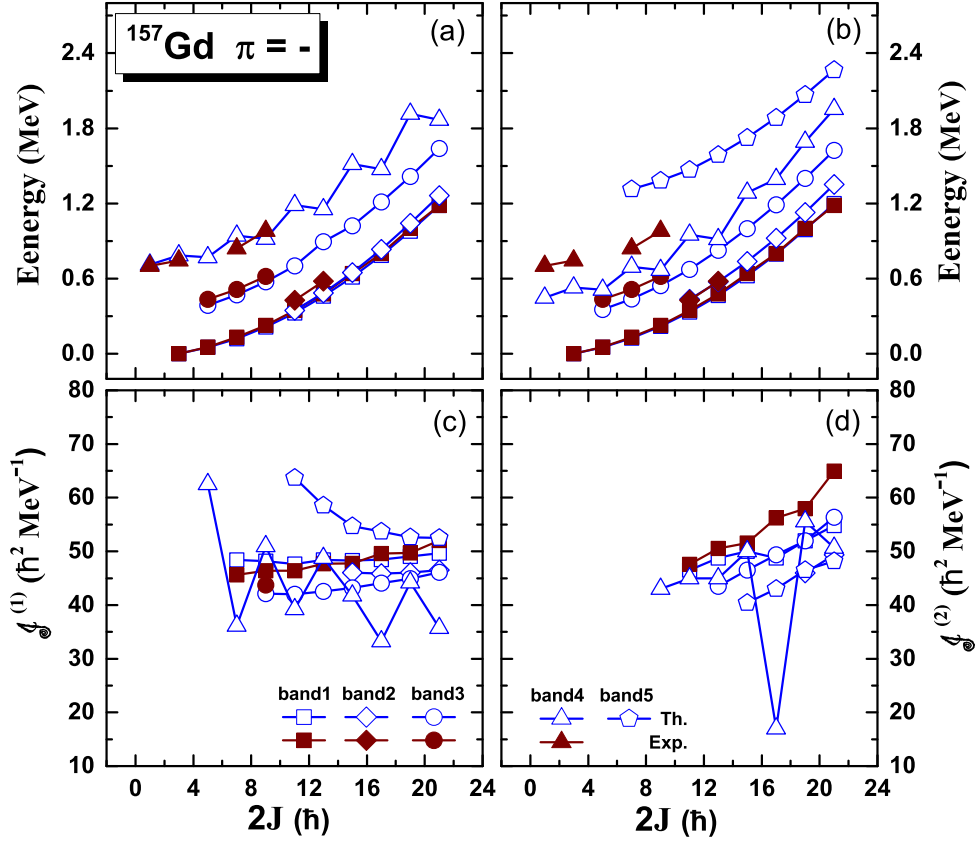}
\caption{(Color online) Same as Fig. \ref{fig:spec-Tb159-po} but for the negative-parity bands of  the odd-neutron nucleus $^{157}$Gd. The Fermi surface and coupling strength ($\lambda$, $\chi$) in the CQC model are chosen as  (-9.70 MeV, 12.50 MeV/$b^2$). }
\label{fig:spec-Gd157-Ne}
\end{figure*}

\begin{figure*}[htb]
\includegraphics[scale=0.5]{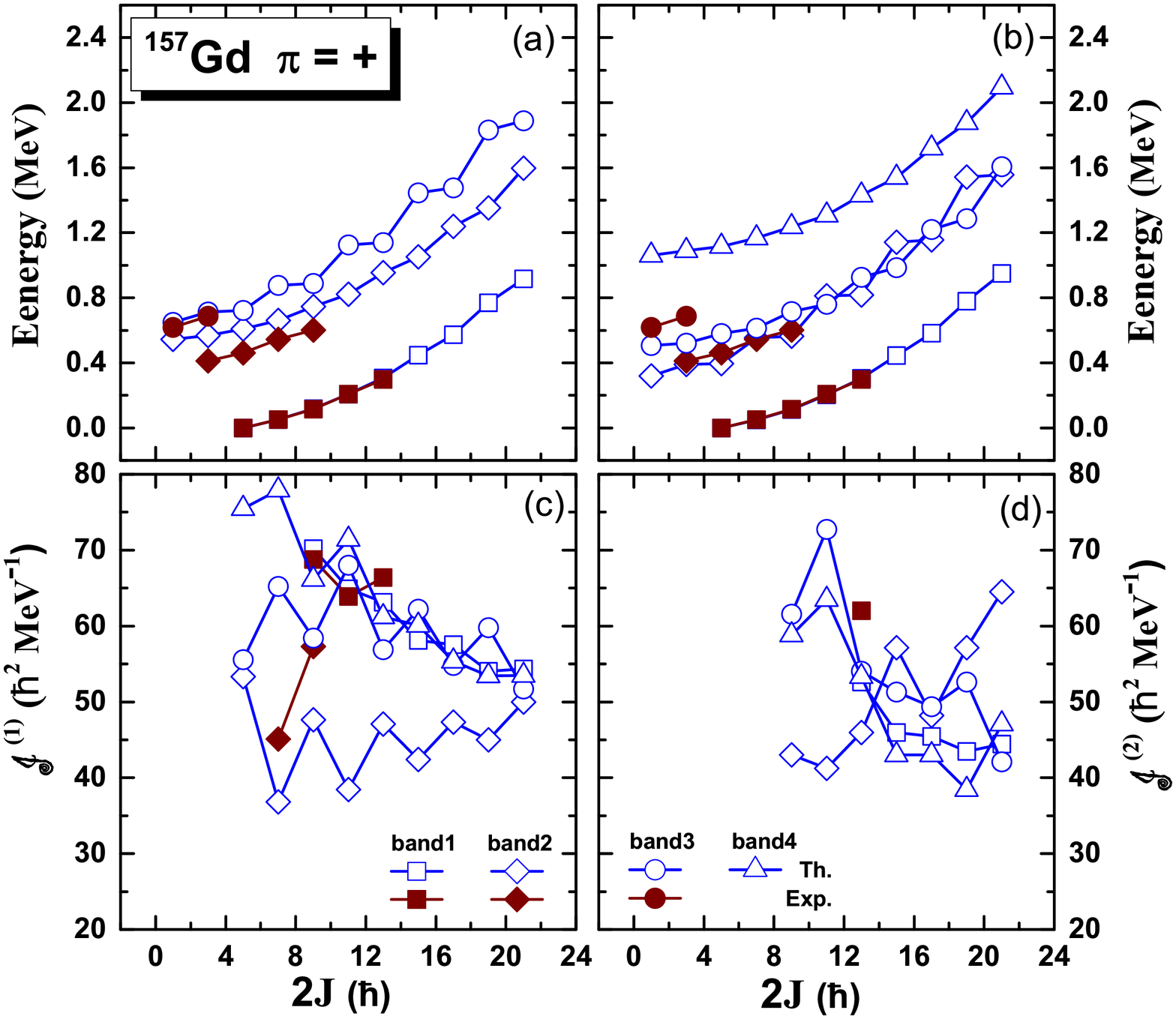}
\caption{(Color online)  Same as Fig. \ref{fig:spec-Tb159-po} but for the positive-parity bands of  $^{157}$Gd. The excitation energies in the upper panels are shown relative to the lowest state.  The Fermi surface and coupling strength ($\lambda$, $\chi$) in the CQC model are chosen as  (-7.70 MeV, 15.10 MeV/$b^2$). }
\label{fig:spec-Gd157-po}
\end{figure*}

Figure \ref{fig:spec-Tb159-po} compares the calculated low-lying positive-parity bands (panels a, b), kinematic moments of inertia ${\mathscr{J}}^{(1)}=\frac{2J-1}{E_\gamma(J)}$ (where $E_\gamma(J)=E(J)-E(J-2)$) (panel c), and dynamic moments of inertia ${\mathscr{J}}^{(2)}=\frac{4}{E_\gamma(J)-E_\gamma(J-2)}$ (panel d) of the odd-proton nucleus $^{159}$Tb to available data~\cite{NNDC}. The theoretical results are calculated from microscopic CQC model with only the ground state band of the core (panel a) and with both the ground state and $\gamma$ bands of the core (panels b, c, d). For the single nucleon valence space in the CQC model, we include the spherical single-particle states located in between $E_f\pm 2\hbar\omega$, where $E_f$ is the Fermi surface of the corresponding spherical configuration and $\hbar\omega=41A^{-1/3}$ MeV. For the $A\sim160$ mass region, states within approximately 15 MeV of the Fermi surface are included, and this is sufficient to calculate the low-lying spectrum.

The levels are grouped into different bands according to the dominant decay pattern.
Here, four lowest-lying measured bands with band heads $J^{\pi}=3/2^{+}$ (0.000 MeV), $J^{\pi}=5/2^{+}$ (0.348 MeV), $J^{\pi}=1/2^{+}$ (0.581 MeV), and $J^{\pi}=7/2^{+}$ (0.777 MeV) are shown. Bands 1, 2, and 3 exhibit strong-coupling $\Delta J=1$ systematics. In Fig. \ref{fig:spec-Tb159-po} (a), the CQC model with ground state band of the core can reproduce most of the structure of the bands in $^{159}$Tb and only the band 3 is $\sim0.2$ MeV higher than the data. Band 1 predominantly corresponds to the $2d_{5/2}$ spherical proton configuration, while bands 2 and 4 are dominated by the $1g_{7/2}$ hole and particle configurations, respectively. Band 3 is based on two strongly mixed configurations of $2d_{3/2}$ and $2d_{5/2}$ spherical single-proton states.
When adding the $\gamma$ band of the core to the CQC model, in Fig. \ref{fig:spec-Tb159-po} (b) bands 1, 2, and 4 have been modified only slightly, while band 3 is lowered by $\sim0.15$ MeV and closer to the data. This is because band 3 has strong mixing between the configurations based on the ground state band ($\sim60$\%) and $\gamma$ band ($\sim40$\%) of the core. The off-diagonal matrix elements of the CQC Hamiltonian mainly come from the configurations with $\Delta R=1$ (e.g. $3_\gamma^+ \leftrightarrow 2^+_{\rm g.s.}$, $3_\gamma^+ \leftrightarrow 4^+_{\rm g.s.}$, $5_\gamma^+ \leftrightarrow 4^+_{\rm g.s.}$, $\cdots$)  and $\Delta j=1$ ($2d_{3/2}\leftrightarrow 2d_{5/2}$), and the typical value is  $\sim 0.15$ MeV, consistent with the shifting of band 3. In Fig. \ref{fig:spec-Tb159-po} (b), a band for which purely coupling to the $\gamma$ band of the core is also plotted as band 5, and the band head is 1.058 MeV, which is almost identical to the calculated $\gamma$ band head, 1.035 MeV, of the core $^{158}$Gd. It is remarkable that the band structure of band 5, e.g. quasiparticle configurations, moments of inertia [c.f. Fig. \ref{fig:spec-Tb159-po} (c, d)], and electromagnetic transitions (c.f. Table \ref{tableTb159-po}), is very similar as that of the ground state band. A possible  candidate of this band head is the measured state with $J^\pi=(7/2^+)$ at 1.102 MeV \cite{NNDC}.

The theoretical and experimental kinematic moments of inertia of the ground state band are $40\sim50$ $\hbar^2\ {\rm MeV}^{-1}$ with the discrepancy between them less than 3 $\hbar^2\ {\rm MeV}^{-1}$, and both increase gradually when moving to high spin. $\mathscr{J}^{(1)}$ of the excitation bands are close to those of the ground state band except for a lower lying band 4. Moreover, $\mathscr{J}^{(1)}$ of band 3 exhibits a staggering behavior but the phase of our prediction is opposite with the data. The dynamic moment of inertia $\mathscr{J}^{(2)}$ is a very sensitive quantity as it describes the variation of $\mathscr{J}^{(1)}$. The calculated $\mathscr{J}^{(2)}$ are in rather good agreement with the data, and both increase more rapidly than the $\mathscr{J}^{(1)}$ with increasing angular momentum. The description of moments of inertia for the ground state band before band crossing using our microscopic CQC model is similar to that by the cranked relativistic Hartree-Bogoliubov \cite{VALR.05} or cranked nonrelativistic Hartree-Fock-Bogoliubov \cite{Valor00}.

In Fig. \ref{fig:spec-Tb159-Ne}, the sequences of the negative-parity levels of $^{159}$Tb built on the states $J^\pi=5/2^-_1$ and $J^\pi=7/2^-_2$ form $\Delta J=1$ rotational bands, and both are originated from the  $1h_{11/2}$ proton configuration. Bands 1 and 2 are almost not changed by including the $\gamma$ band of the core. In Fig. \ref{fig:spec-Tb159-Ne} (b), we also plot the calculated lowest excited states based on the $\gamma$ band of the core (open circles). There are no strong cascaded electromagnetic transitions between the states, and therefore they are not denoted as a band. The measured states $J^\pi=(1/2^-)$ at 0.855 MeV and $J^\pi=(5/2^-)$ at 0.891 MeV are possible candidates for the $\gamma$ phonon excitation states. The calculated moments of inertia $\mathscr{J}^{(1)}$ and $\mathscr{J}^{(2)}$ are all in good agreement with the data. The moments of inertia of the two bands are rather different at low spin but agree better at high spin. This may be because the Coriolis coupling of band 1 is larger than that of band 2 at low spin and becomes similar for high spin states \cite{Ring80}.

Figure \ref{fig:spec-Gd157-Ne} displays the low-lying negative-parity bands and corresponding moments of inertia for the odd-neutron nucleus $^{157}$Gd.
The lowest three bands have been reproduced quite well by CQC model without and with the $\gamma$ band of the core. Bands 1, 2, and 3 predominately correspond to the strongly mixed configurations of $2f_{7/2}$ and $1h_{9/2}$, a rather pure $1h_{11/2}$ configuration, and a rather pure $1h_{9/2}$ configuration, respectively. Similar as the band 3 in Fig. \ref{fig:spec-Tb159-po}, the band 4 here built on $J^\pi=1/2^-_1$ is shifted by $\sim0.25$ MeV when the $\gamma$ band of the core is included and lower than the data. It is also noted that the calculated band 4 presents a staggering possibly due to a Coriolis coupling that is too strong, and this could be solved by adding a magnetic dipole particle-core interaction term to the present model \cite{Prot97}.
In Fig. \ref{fig:spec-Gd157-Ne} (b), band 5 is a $\gamma$ phonon excitation band built on $J^\pi=7/2^-_5$ at 1.313 MeV, close to the calculated $\gamma$ band head, 1.152 MeV, of the core $^{156}$Gd. Moreover, the dominated configurations of band 5 are similar as those of band 1.
The CQC model can reproduce the moments of inertia for the ground state band of $^{157}$Gd very well. The quasiparticle excitation bands 2 and 3 share similar moments of inertia with those of band 1. The signature splitting of calculated band 4 results in staggered moments of inertia, while band 5 has a decreasing $\mathscr{J}^{(1)}$ and an increasing $\mathscr{J}^{(2)}$ as functions of spin.

For the positive-parity bands of $^{157}$Gd in Fig. \ref{fig:spec-Gd157-po}, the calculated band 1, originated from the  $1i_{13/2}$ neutron configuration, is in good agreement with the data for both excitation energies and $\mathscr{J}^{(1)}$. When the $\gamma$ band of the core is included, band 3 built on $J^\pi=1/2^+_{2}$ in Fig. \ref{fig:spec-Gd157-po} (a)  is lowered by $\sim0.3$ MeV and denoted as band 2 in Fig. \ref{fig:spec-Gd157-po} (b) since they have the same dominant single-particle configurations. Then the theoretical results reproduce the data for the excitation energies and the trend of $\mathscr{J}^{(1)}$ of band 2. Band 3 in Fig. \ref{fig:spec-Gd157-po} (b) corresponds to band 2 in Fig. \ref{fig:spec-Gd157-po} (a), and is almost unchanged by including $\gamma$ band.  Band 4 is a  $\gamma$ phonon excitation band and possesses similar single-particle configurations and moments of inertia as those of band 1.

\begin{figure}[htb]
\includegraphics[scale=0.22]{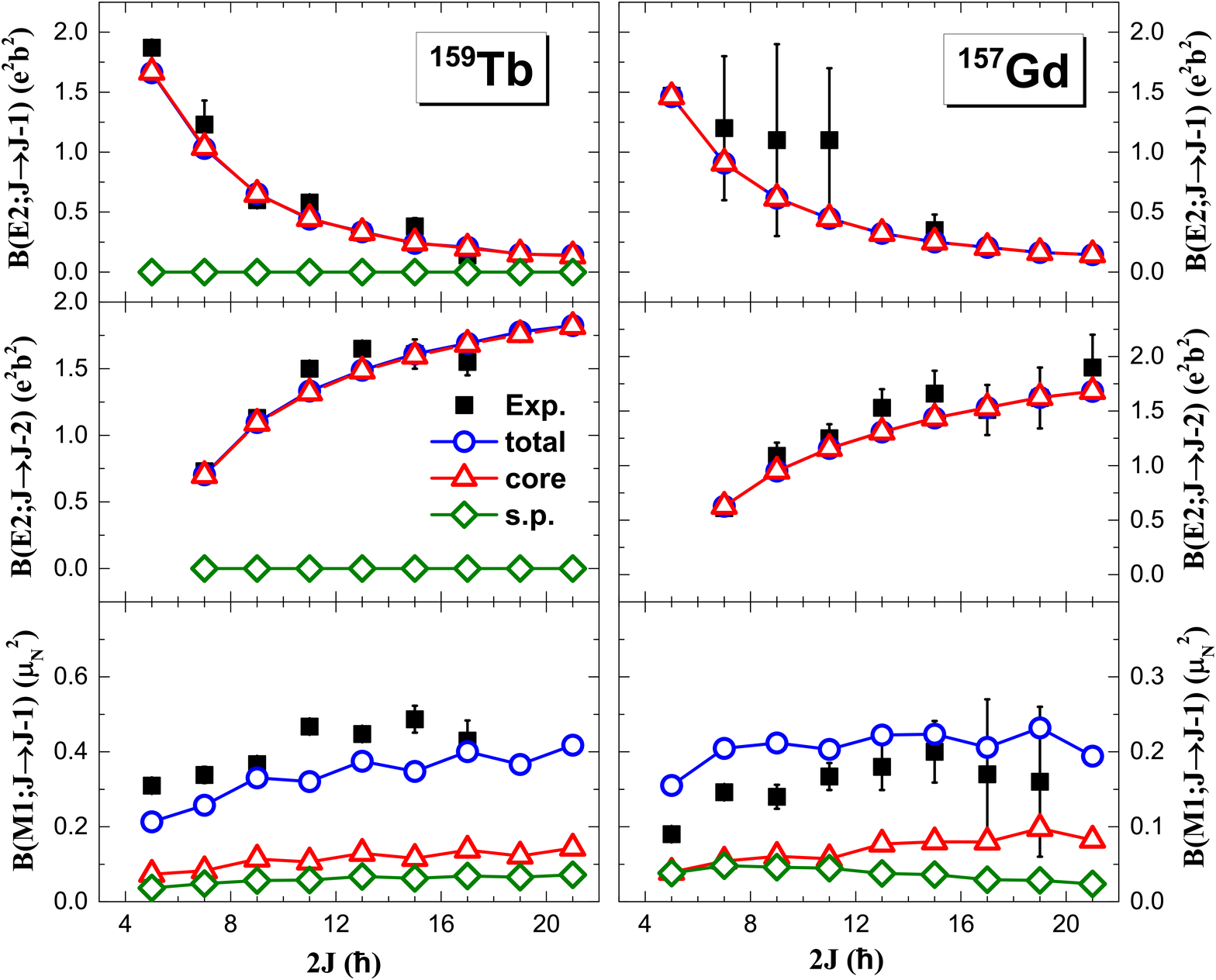}
\caption{(Color online)
The core and single-particle contributions to the intraband $B(E2; J\to J-1)$, $B(E2; J\to J-2)$,
and $B(M1; J\to J-1)$ in the ground state bands  of $^{159}$Tb (left panels) and $^{157}$Gd (right panels). In $^{157}$Gd, the single neutron does not contribute to the $E2$ transitions.}
\label{fig:Tb159EM2}
\end{figure}

Tables \ref{tableTb159-po}, \ref{tableTb159-Ne}, \ref{tableGd157-Ne}, and \ref{tableGd157-po} collect the results for intraband and interband electric quadrupole $E2$ and magnetic dipole $M1$ transition rates of $^{159}$Tb and $^{157}$Gd. The theoretical results are calculated from the microscopic CQC model with both the ground state and $\gamma$ bands of the core. For the $E2$ transition, the bare charge of a proton is used. For the $M1$ transition, three $g$-factors are necessary: $g_c$ for the core; and $g_s$ and $g_l$ for the spin and orbital parts of single particle, respectively. In the present work, $g_c=Z/A$ is used for the well-deformed even-mass core and $g_l=1 (0)$ for the single proton (neutron). The spin $g$-factor $g_s$ is quenched by 30\% with respect to the value of the free nucleon in this mass region to simulate the spin polarization effect. This polarization effect can be described in terms of the coupling to excitations of the even core produced by spin-dependent fields, which are associated with the presence of unsaturated spins \cite{Bohr75}. The theoretical results are in very good agreement with the data for intraband $E2$ transitions of the ground state band in $^{159}$Tb and $^{157}$Gd. The model also reproduces the systematic trend of the $M1$ transitions but fails in the description of the staggering behavior ({\it c.f.}, Fig. \ref{fig:Tb159EM2}). This can be understood from the wave functions in Tab. \ref{tablewf}, where the dominant configurations for some selected states of the ground state band in $^{159}$Tb are listed as examples. All the states predominately correspond to the  $2d_{5/2}$ single-particle configuration. $J^\pi=11/2^+$ and $13/2^+$ have larger overlap  and consequently larger $M1$ matrix element according to Eqs. (\ref{eq:MUsp}-\ref{eq:MVc}, \ref{eq:M1c}) than that between $J^\pi=11/2^+$ and $9/2^+$. This leads to a stronger $B(M1; 13/2^+\to 11/2^+)$ than $B(M1; 11/2^+\to 9/2^+)$. Similar results are also found in the states $J^\pi=13/2^+$, $15/2^+$, and $17/2^+$.

There are no available experimental data for the intraband and interband transitions of the excitation bands. The theoretical results for intraband $E2$ transitions of the excitation bands have similar trends and quantitates as those of the ground state bands except for the positive-parity $\gamma$ phonon excitation band (band 4) in $^{157}$Gd, while the intraband $B(M1)$ are rather different since they are sensitive to the dominant single-particle configurations of the bands. For the interband transitions, we only list relatively larger transitions and they are generally much smaller than those of the intraband transitions. However, the interband transitions for the positive-parity band 4 to band 2 and negative-parity band 2 to band 1 in $^{159}$Tb are rather large because the connected two bands share similar single-particle configurations.

\begin{center}
\small
\topcaption{The calculated intraband and interband $E2$ (in units of $e^{2}b^{2}$) and $M1$ (in units of $\mu_{N}^{2}$)
transition rates for low-lying positive-parity bands in $^{159}$Tb, compared to available data~\cite{Reich12b}.}
\tablefirsthead{\hline  \hline\multicolumn{1}{c|}{}&
\multicolumn{1}{c}{} &
   \multicolumn{1}{c}{Th.} &
    \multicolumn{1}{c}{Exp.} &
     \multicolumn{1}{c}{}&
     \multicolumn{1}{c}{Th.} &
      \multicolumn{1}{c}{Exp.} \\\hline}

\tablehead{\multicolumn{3}{l}{continue}\\\hline\hline
\multicolumn{1}{c|}{} &
\multicolumn{1}{c}{} &
   \multicolumn{1}{c}{Th.} &
    \multicolumn{1}{c}{Exp.} &
     \multicolumn{1}{c}{}&
     \multicolumn{1}{c}{Th.} &
      \multicolumn{1}{c}{Exp.} \\\hline}

\tabletail{\hline\multicolumn{3}{r}{\small }\\}

\begin{supertabular}{p{25mm}|p{42mm}p{11mm}<{\centering}p{15mm}<{\centering}p{41mm}p{11mm}<{\centering}p{15mm}<{\centering}}
\footnotesize band1 $\rightarrow$ band1
&$B(E2;7/2_{1}^{+}\rightarrow 3/2_{1}^{+})$      & 0.70    & 0.73(4)  &$B(E2;5/2_{1}^{+}\rightarrow 3/2_{1}^{+})$ &1.66&1.87(5)\\

&$B(E2;9/2_{1}^{+}\rightarrow 5/2_{1}^{+})$      & 1.10    & 1.13(5) &$B(E2;7/2_{1}^{+}\rightarrow 5/2_{1}^{+})$  &1.03&1.23(20)\\
&$B(E2;11/2_{1}^{+}\rightarrow 7/2_{1}^{+})$     & 1.33   & 1.50(3)  &$B(E2;9/2_{1}^{+}\rightarrow 7/2_{1}^{+})$ &0.65 &0.60(6)\\

&$B(E2;13/2_{1}^{+}\rightarrow 9/2_{1}^{+})$     & 1.49    & 1.65(5) &$B(E2;11/2_{1}^{+}\rightarrow 9/2_{1}^{+})$ & 0.44&0.58(6)\\
&$B(E2;15/2_{1}^{+}\rightarrow 11/2_{1}^{+})$    & 1.61   & 1.61(11) &$B(E2;13/2_{1}^{+}\rightarrow 11/2_{1}^{+})$&0.34&0.33(4)\\

&$B(E2;17/2_{1}^{+}\rightarrow 13/2_{1}^{+})$    & 1.69  & 1.55(10) &$B(E2;15/2_{1}^{+}\rightarrow 13/2_{1}^{+})$& 0.24&0.38(7)\\

&$B(E2;19/2_{1}^{+}\rightarrow 15/2_{1}^{+})$    & 1.78   &  -  & $B(E2;17/2_{1}^{+}\rightarrow 15/2_{1}^{+})$& 0.21 & 0.14(7)\\

&$B(E2;21/2_{1}^{+}\rightarrow 17/2_{1}^{+})$     & 1.82  & 2.00(26)  &$B(E2;19/2_{1}^{+}\rightarrow 17/2_{1}^{+})$& 0.15  & - \\

&$B(M1;5/2_{1}^{+}\rightarrow 3/2_{1}^{+})$     & 0.21   & 0.310(14)  & $B(E2;21/2_{1}^{+}\rightarrow 19/2_{1}^{+})$ &0.14  &-\\

&$B(M1;7/2_{1}^{+}\rightarrow 5/2_{1}^{+})$ & 0.26   & 0.338(21)& $B(M1;15/2_{1}^{+}\rightarrow 13/2_{1}^{+})$&0.35&0.487(36)\\

&$B(M1;9/2_{1}^{+}\rightarrow 7/2_{1}^{+})$ & 0.33   & 0.367(18) &$B(M1;17/2_{1}^{+}\rightarrow 15/2_{1}^{+})$&0.40&0.430(54)\\

&$B(M1;11/2_{1}^{+}\rightarrow 9/2_{1}^{+})$   & 0.32   & 0.467(9)  &$B(M1;19/2_{1}^{+}\rightarrow 17/2_{1}^{+})$&0.37   & -\\

&$B(M1;13/2_{1}^{+}\rightarrow 11/2_{1}^{+})$ & 0.37 &0.448(18)  &  $B(M1;21/2_{1}^{+}\rightarrow 19/2_{1}^{+})$ &0.42   & -\\
 \hline
\footnotesize band2 $\rightarrow$ band2
&$B(E2;9/2_{2}^{+}\rightarrow 5/2_{2}^{+})$      & 0.53   & -  &$B(E2;7/2_{2}^{+}\rightarrow 5/2_{2}^{+})$      &1.88  & -\\

&$B(E2;11/2_{2}^{+}\rightarrow 7/2_{2}^{+})$      & 0.92   & -  &$B(E2;9/2_{2}^{+}\rightarrow 7/2_{2}^{+})$      &1.61   & -\\

&$B(E2;13/2_{2}^{+}\rightarrow 9/2_{2}^{+})$      & 1.18   & -  &$B(E2;11/2_{2}^{+}\rightarrow 9/2_{2}^{+})$      &1.26  & -\\

&$B(E2;15/2_{2}^{+}\rightarrow 11/2_{2}^{+})$      & 1.38  & -  &$B(E2;13/2_{2}^{+}\rightarrow 11/2_{2}^{+})$      &0.99   &-\\

&$B(M1;7/2_{2}^{+}\rightarrow 5/2_{2}^{+})$      & 1.18   & - &$B(E2;15/2_{2}^{+}\rightarrow 13/2_{2}^{+})$      &0.78   & -\\

&$B(M1;9/2_{2}^{+}\rightarrow 7/2_{2}^{+})$      & 1.77  & -   &$B(M1;13/2_{2}^{+}\rightarrow 11/2_{2}^{+})$      &2.34   &-\\

&$B(M1;11/2_{2}^{+}\rightarrow 9/2_{2}^{+})$      & 2.14   & - &$B(M1;15/2_{2}^{+}\rightarrow 13/2_{2}^{+})$      &2.55   &-\\
\footnotesize band2 $\rightarrow$ band1
&$B(M1;7/2_{2}^{+}\rightarrow 5/2_{1}^{+})$      &0.09 & -&$B(M1;13/2_{2}^{+}\rightarrow 11/2_{1}^{+})$      &0.04 & -\\

&$B(M1;9/2_{2}^{+}\rightarrow 7/2_{1}^{+})$      &0.07  & - &      & & \\

\hline
\footnotesize band3 $\rightarrow$ band3
&$B(E2;5/2_{3}^{+}\rightarrow 1/2_{1}^{+})$      & 0.34  & -  &$B(E2;3/2_{2}^{+}\rightarrow 1/2_{1}^{+})$      &0.12   & -\\

&$B(E2;7/2_{4}^{+}\rightarrow 3/2_{2}^{+})$      & 0.64   & -  &$B(E2;5/2_{3}^{+}\rightarrow 3/2_{2}^{+})$      &0.02  & -\\

&$B(E2;9/2_{4}^{+}\rightarrow 5/2_{3}^{+})$      & 0.73 & - &  $B(E2;9/2_{4}^{+}\rightarrow 7/2_{4}^{+})$      &0.03  & -\\

&$B(E2;11/2_{3}^{+}\rightarrow 7/2_{4}^{+})$      & 1.06  & - &$B(E2;11/2_{3}^{+}\rightarrow 9/2_{4}^{+})$      &0.04 & - \\

&$B(M1;3/2_{2}^{+}\rightarrow 1/2_{1}^{+})$      & 0.23  & - &$B(M1;7/2_{4}^{+}\rightarrow 5/2_{3}^{+})$      &0.15  & -  \\

&$B(M1;5/2_{3}^{+}\rightarrow 3/2_{2}^{+})$      & 0.09  & -  &$B(M1;11/2_{3}^{+}\rightarrow 9/2_{4}^{+})$      &0.14  & - \\

\footnotesize band3 $\rightarrow$ band1
&$B(E2;11/2_{3}^{+}\rightarrow 9/2_{1}^{+})$      & 0.01  & -  &$B(M1;5/2_{3}^{+}\rightarrow 3/2_{1}^{+})$      &0.03 & -\\
\hline
\footnotesize band4 $\rightarrow$ band4&$B(E2;11/2_{4}^{+}\rightarrow 7/2_{3}^{+})$      & 0.37   & -  &$B(E2;9/2_{3}^{+}\rightarrow 7/2_{3}^{+})$      &1.82   & -\\
&$B(E2;13/2_{4}^{+}\rightarrow 9/2_{3}^{+})$      & 0.70  & -  &$B(E2;11/2_{4}^{+}\rightarrow 9/2_{3}^{+})$      &1.81   & -\\

&$B(E2;15/2_{4}^{+}\rightarrow 11/2_{4}^{+})$      & 0.95   & -  &$B(E2;13/2_{4}^{+}\rightarrow 11/2_{4}^{+})$      &1.52  &-\\

&$B(E2;17/2_{4}^{+}\rightarrow 13/2_{4}^{+})$      & 1.15   & -  &$B(E2;15/2_{4}^{+}\rightarrow 13/2_{4}^{+})$      &1.23  & -\\

&$B(M1;9/2_{3}^{+}\rightarrow 7/2_{3}^{+})$      & 2.67   & -  &$B(E2;17/2_{4}^{+}\rightarrow 15/2_{4}^{+})$      &0.98  & -\\

&$B(M1;11/2_{4}^{+}\rightarrow 9/2_{3}^{+})$      & 4.05   & -  &$B(M1;15/2_{4}^{+}\rightarrow 13/2_{4}^{+})$      &5.31  & -\\

&$B(M1;13/2_{4}^{+}\rightarrow 11/2_{4}^{+})$      & 4.84  & -  &$B(M1;17/2_{4}^{+}\rightarrow 15/2_{4}^{+})$      &5.56  & -\\
\footnotesize band4 $\rightarrow$ band2
&$B(E2;7/2_{3}^{+}\rightarrow 5/2_{2}^{+})$      & 0.03   & -  &$B(E2;9/2_{3}^{+}\rightarrow 5/2_{2}^{+})$      &0.01 & -\\

&$B(M1;7/2_{3}^{+}\rightarrow 5/2_{2}^{+})$      & 0.94  & -  &$B(M1;11/2_{4}^{+}\rightarrow 9/2_{2}^{+})$      & 0.45  & - \\

&$B(M1;9/2_{3}^{+}\rightarrow 7/2_{2}^{+})$      &0.62 & -  &$B(M1;13/2_{4}^{+}\rightarrow 11/2_{2}^{+})$      &0.33 & -\\

\hline
\footnotesize band5 $\rightarrow$ band5
&$B(E2;11/2_{5}^{+}\rightarrow 7/2_{5}^{+})$      & 0.20   & -  &$B(E2;9/2_{5}^{+}\rightarrow 7/2_{5}^{+})$      &0.08  & -\\

&$B(E2;13/2_{5}^{+}\rightarrow 9/2_{5}^{+})$      & 0.86  & -  &$B(E2;11/2_{5}^{+}\rightarrow 9/2_{5}^{+})$      &0.05  & -\\

&$B(E2;15/2_{5}^{+}\rightarrow 11/2_{5}^{+})$      & 1.29   & -  &$B(E2;13/2_{5}^{+}\rightarrow 11/2_{5}^{+})$    &0.08  & - \\

&$B(E2;17/2_{5}^{+}\rightarrow 13/2_{5}^{+})$      & 1.53   & -  &$B(E2;15/2_{5}^{+}\rightarrow 13/2_{5}^{+})$    &0.06  & -  \\

&$B(M1;9/2_{5}^{+}\rightarrow 7/2_{5}^{+})$      & 0.18  & -  &$B(M1;13/2_{5}^{+}\rightarrow 11/2_{5}^{+})$    &0.34  & - \\

&$B(M1;11/2_{5}^{+}\rightarrow 9/2_{5}^{+})$      & 0.27   & -  &$B(M1;15/2_{5}^{+}\rightarrow 13/2_{5}^{+})$    &0.39  & -  \\

\footnotesize band5 $\rightarrow$ band3
&$B(E2;7/2_{5}^{+}\rightarrow 5/2_{3}^{+})$      & 0.02  & -  &$B(E2;9/2_{5}^{+}\rightarrow 7/2_{4}^{+})$      &0.03 & -\\

\footnotesize band5 $\rightarrow$ band4
&$B(E2;15/2_{5}^{+}\rightarrow 13/2_{4}^{+})$      & 0.05  & -  &$B(E2;17/2_{5}^{+}\rightarrow 15/2_{4}^{+})$      &0.06  & -\\
\hline
\hline
\end{supertabular}
\label{tableTb159-po}
\end{center}

\begin{center}
\small
\topcaption{The calculated intraband and interband $E2$ (in units of $e^{2}b^{2}$) and $M1$ (in units of $\mu_{N}^{2}$)
transition rates for low-lying negative-parity bands in $^{159}$Tb.}
\tablefirsthead{\hline  \hline\multicolumn{1}{c}{}&
\multicolumn{1}{c}{} &
   \multicolumn{1}{c}{Th.} &
    \multicolumn{1}{c}{Exp.} &
     \multicolumn{1}{c}{}&
     \multicolumn{1}{c}{Th.} &
      \multicolumn{1}{c}{Exp.} \\\hline}

\tablehead{\multicolumn{3}{l}{continue}\\\hline\hline
\multicolumn{1}{c|}{} &
\multicolumn{1}{c}{} &
   \multicolumn{1}{c}{Th.} &
    \multicolumn{1}{c}{Exp.} &
     \multicolumn{1}{c}{}&
     \multicolumn{1}{c}{Th.} &
      \multicolumn{1}{c}{Exp.} \\\hline}
\tabletail{\hline\multicolumn{3}{r}{\small }\\}

\begin{supertabular}{p{25mm}|p{42mm}<{\centering}p{11mm}<{\centering}p{15mm}<{\centering}p{42mm}p{11mm}<{\centering}p{15mm}<{\centering}}

\footnotesize band1 $\rightarrow$ band1
&$B(E2;9/2_{1}^{+}\rightarrow 5/2_{1}^{+})$ & 0.40 & -  &$B(E2;7/2_{1}^{+}\rightarrow 5/2_{1}^{+})$& 1.37 & -\\

&$B(E2;11/2_{1}^{+}\rightarrow 7/2_{1}^{+})$     & 0.76   & -  &$B(E2;9/2_{1}^{+}\rightarrow 7/2_{1}^{+})$& 1.60 & -\\

&$B(E2;13/2_{1}^{+}\rightarrow 9/2_{1}^{+})$     & 1.06   & - &$B(E2;11/2_{1}^{+}\rightarrow 9/2_{1}^{+})$& 1.36  & - \\

&$B(E2;15/2_{1}^{+}\rightarrow 11/2_{1}^{+})$     & 1.26   & - &$B(E2;13/2_{1}^{+}\rightarrow 11/2_{1}^{+})$& 1.08 & - \\

&$B(M1;9/2_{1}^{+}\rightarrow 7/2_{1}^{+})$ & 0.22 & - &  $B(E2;15/2_{1}^{+}\rightarrow 13/2_{1}^{+})$& 0.87 & -  \\

&$B(M1;11/2_{1}^{+}\rightarrow 9/2_{1}^{+})$ & 0.55   & - &  $B(M1;15/2_{1}^{+}\rightarrow 13/2_{1}^{+})$& 1.10 & -   \\

&$B(M1;13/2_{1}^{+}\rightarrow 11/2_{1}^{+})$& 0.72& - &  &  &   \\
\hline
\footnotesize band2$\rightarrow$ band2
&$B(E2;11/2_{2}^{+}\rightarrow 7/2_{2}^{+})$      & 0.51   & -  &$B(E2;9/2_{2}^{+}\rightarrow 7/2_{2}^{+})$      &1.69   & -\\

&$B(E2;13/2_{2}^{+}\rightarrow 9/2_{2}^{+})$      & 0.88   & -  &$B(E2;11/2_{2}^{+}\rightarrow 9/2_{2}^{+})$      &1.56   & -\\

&$B(E2;15/2_{2}^{+}\rightarrow 11/2_{2}^{+})$      & 1.09   & -  &$B(E2;13/2_{2}^{+}\rightarrow 11/2_{2}^{+})$      &1.27   &-\\

&$B(E2;17/2_{2}^{+}\rightarrow 13/2_{2}^{+})$      & 1.31  & -  &$B(E2;15/2_{2}^{+}\rightarrow 13/2_{2}^{+})$      &1.03   &-\\

&$B(M1;9/2_{2}^{+}\rightarrow 7/2_{2}^{+})$      & 1.85  & -  &  $B(E2;17/2_{2}^{+}\rightarrow 15/2_{2}^{+})$   &0.81   &-\\

&$B(M1;11/2_{2}^{+}\rightarrow 9/2_{2}^{+})$      & 2.34 & -  & $B(M1;15/2_{2}^{+}\rightarrow 13/2_{2}^{+})$     &2.61   &-\\

&$B(M1;13/2_{2}^{+}\rightarrow 11/2_{2}^{+})$      & 2.29   & -  & $B(M1;17/2_{2}^{+}\rightarrow 15/2_{2}^{+})$  &2.06 &-\\
\footnotesize band2 $\rightarrow$ band1
&$B(E2;7/2_{2}^{+}\rightarrow 5/2_{1}^{+})$      & 0.45   & -  &$B(E2;7/2_{2}^{+}\rightarrow 7/2_{1}^{+})$      &0.32   &-\\

&$B(E2;9/2_{2}^{+}\rightarrow 5/2_{1}^{+})$      & 0.16  & -  &$B(E2;9/2_{2}^{+}\rightarrow 9/2_{1}^{+})$      &0.16  &-\\

&$B(M1;7/2_{2}^{+}\rightarrow 5/2_{1}^{+})$      & 2.49  & -  &  $B(M1;11/2_{2}^{+}\rightarrow 9/2_{1}^{+})$   &0.61   &-\\

&$B(M1;9/2_{2}^{+}\rightarrow 7/2_{1}^{+})$      & 0.98  & -  & $B(M1;15/2_{2}^{+}\rightarrow 13/2_{1}^{+})$     &0.45   &-\\
\hline
\hline
\end{supertabular}
\label{tableTb159-Ne}
\end{center}

\begin{center}
\small
\topcaption{The calculated intraband and interband $E2$ (in units of $e^{2}b^{2}$) and $M1$ (in units of $\mu_{N}^{2}$)
transition rates for low-lying negative-parity bands in $^{157}$Gd, compared to available data \cite{Kusa92}.}
\tablefirsthead{\hline  \hline\multicolumn{1}{c}{}&
\multicolumn{1}{c}{} &
   \multicolumn{1}{c}{Th.} &
    \multicolumn{1}{c}{Exp.} &
     \multicolumn{1}{c}{}&
     \multicolumn{1}{c}{Th.} &
      \multicolumn{1}{c}{Exp.} \\\hline}

\tablehead{\multicolumn{3}{l}{continue}\\\hline\hline
\multicolumn{1}{c|}{} &
\multicolumn{1}{c}{} &
   \multicolumn{1}{c}{Th.} &
    \multicolumn{1}{c}{Exp.} &
     \multicolumn{1}{c}{}&
     \multicolumn{1}{c}{Th.} &
      \multicolumn{1}{c}{Exp.} \\\hline}

\tabletail{\hline\multicolumn{3}{r}{\small }\\}

\begin{supertabular}{p{25mm}|p{42mm}p{11mm}<{\centering}p{15mm}<{\centering}p{42mm}p{11mm}<{\centering}p{15mm}<{\centering}}
\footnotesize band1 $\rightarrow$ band1
&$B(E2;7/2_{1}^{+}\rightarrow 3/2_{1}^{+})$ & 0.63& 0.61(5) &$B(E2;5/2_{1}^{+}\rightarrow 3/2_{1}^{+})$& 1.46    &1.47(7)\\

&$B(E2;9/2_{1}^{+}\rightarrow 5/2_{1}^{+})$ & 0.95 & 1.09(12) &$B(E2;7/2_{1}^{+}\rightarrow 5/2_{1}^{+})$& 0.91&1.20(60)\\

&$B(E2;11/2_{1}^{+}\rightarrow 7/2_{1}^{+})$ & 1.16  & 1.25(13) &$B(E2;9/2_{1}^{+}\rightarrow 7/2_{1}^{+})$&   0.62& 1.10(80)\\

&$B(E2;13/2_{1}^{+}\rightarrow 9/2_{1}^{+})$ & 1.31 &1.53(17) &$B(E2;11/2_{1}^{+}\rightarrow 9/2_{1}^{+})$& 0.45& 1.10(60)  \\

&$B(E2;15/2_{1}^{+}\rightarrow 11/2_{1}^{+})$ & 1.44 &1.66(21) &$B(E2;13/2_{1}^{+}\rightarrow 11/2_{1}^{+})$& 0.32& -  \\

&$B(E2;17/2_{1}^{+}\rightarrow 13/2_{1}^{+})$ & 1.53 &1.51(23) &$B(E2;15/2_{1}^{+}\rightarrow 13/2_{1}^{+})$& 0.25& 0.35(13)  \\

&$B(E2;19/2_{1}^{+}\rightarrow 15/2_{1}^{+})$ & 1.62 &1.62(28) &$B(E2;17/2_{1}^{+}\rightarrow 15/2_{1}^{+})$& 0.21& -  \\

&$B(E2;21/2_{1}^{+}\rightarrow 17/2_{1}^{+})$ & 1.68 &1.90(30) &$B(E2;19/2_{1}^{+}\rightarrow 17/2_{1}^{+})$& 0.16& -  \\

&$B(M1;5/2_{1}^{+}\rightarrow 3/2_{1}^{+})$    & 0.16  & 0.090(7) &  $B(E2;21/2_{1}^{+}\rightarrow 19/2_{1}^{+})$& 0.15& -  \\

&$B(M1;7/2_{1}^{+}\rightarrow 5/2_{1}^{+})$  & 0.20& 0.146(8)& $B(M1;15/2_{1}^{+}\rightarrow 13/2_{1}^{+})$& 0.22& 0.200(40)  \\

&$B(M1;9/2_{1}^{+}\rightarrow 7/2_{1}^{+})$ & 0.21& 0.140(16) &$B(M1;17/2_{1}^{+}\rightarrow 15/2_{1}^{+})$& 0.21 & 0.170(100)\\

&$B(M1;11/2_{1}^{+}\rightarrow 9/2_{1}^{+})$ & 0.20& 0.167(18) &$B(M1;19/2_{1}^{+}\rightarrow 17/2_{1}^{+})$& 0.23 & 0.160(100)\\

&$B(M1;13/2_{1}^{+}\rightarrow 11/2_{1}^{+})$ & 0.22& 0.180(31) &$B(M1;21/2_{1}^{+}\rightarrow 19/2_{1}^{+})$& 0.19 & -\\

\hline
\footnotesize band2 $\rightarrow$ band2
&$B(E2;15/2_{2}^{+}\rightarrow 11/2_{2}^{+})$      & 0.53  & -  &$B(E2;13/2_{2}^{+}\rightarrow 11/2_{2}^{+})$      &1.81   & -\\

&$B(E2;17/2_{2}^{+}\rightarrow 13/2_{2}^{+})$      & 0.78   & -  &$B(E2;15/2_{2}^{+}\rightarrow 13/2_{2}^{+})$      &1.72   & -\\

&$B(E2;19/2_{2}^{+}\rightarrow 15/2_{2}^{+})$      & 0.98  & -  &$B(E2;17/2_{2}^{+}\rightarrow 15/2_{2}^{+})$      &1.53  &-\\

&$B(E2;21/2_{2}^{+}\rightarrow 17/2_{2}^{+})$      & 1.15   & -  &$B(E2;19/2_{2}^{+}\rightarrow 17/2_{2}^{+})$      &1.33  &-\\

&$B(M1;13/2_{2}^{+}\rightarrow 11/2_{2}^{+})$      & 0.75   & -  &$B(E2;21/2_{2}^{+}\rightarrow 19/2_{2}^{+})$      &1.16   &-\\

&$B(M1;15/2_{2}^{+}\rightarrow 13/2_{2}^{+})$      & 1.04   & -  &$B(M1;19/2_{2}^{+}\rightarrow 17/2_{2}^{+})$      &1.42  &-\\

&$B(M1;17/2_{2}^{+}\rightarrow 15/2_{2}^{+})$      & 1.26  & -  &$B(M1;21/2_{2}^{+}\rightarrow 19/2_{2}^{+})$     &1.54  &-\\
\hline
\footnotesize band3 $\rightarrow$ band3
&$B(E2;9/2_{3}^{+}\rightarrow 5/2_{2}^{+})$      & 0.48   & -  &$B(E2;7/2_{2}^{+}\rightarrow 5/2_{2}^{+})$      &1.55   & -\\

&$B(E2;11/2_{4}^{+}\rightarrow 7/2_{2}^{+})$      & 0.82  & -  &$B(E2;9/2_{3}^{+}\rightarrow 7/2_{2}^{+})$      &1.31  &-\\

&$B(E2;13/2_{3}^{+}\rightarrow 9/2_{3}^{+})$      & 1.08   & -  &$B(E2;11/2_{4}^{+}\rightarrow 9/2_{3}^{+})$      &0.98   &-\\

&$B(E2;15/2_{3}^{+}\rightarrow 11/2_{4}^{+})$      & 1.26   & -  &$B(E2;13/2_{3}^{+}\rightarrow 11/2_{4}^{+})$      &0.72  &-\\

&$B(M1;7/2_{2}^{+}\rightarrow 5/2_{2}^{+})$    & 0.02   & -  &$B(E2;15/2_{3}^{+}\rightarrow 13/2_{3}^{+})$      &0.56  &-\\

&$B(M1;13/2_{3}^{+}\rightarrow 11/2_{4}^{+})$      &0.06  &- &$B(M1;15/2_{3}^{+}\rightarrow 13/2_{3}^{+})$      &0.05  &-\\

\footnotesize band3 $\rightarrow$ band1
&$B(E2;5/2_{2}^{+}\rightarrow 3/2_{1}^{+})$   & 0.02 & -  &$B(E2;7/2_{2}^{+}\rightarrow 3/2_{1}^{+})$      &0.03  &-\\

&$B(M1;5/2_{2}^{+}\rightarrow 3/2_{1}^{+})$   & 0.03 & -  &  & & \\

\hline
\footnotesize band4 $\rightarrow$ band4
&$B(E2;5/2_{3}^{+}\rightarrow 1/2_{1}^{+})$      & 0.50   & -  &$B(E2;3/2_{2}^{+}\rightarrow 1/2_{1}^{+})$      &0.51   & -\\

&$B(E2;7/2_{3}^{+}\rightarrow 3/2_{2}^{+})$      & 0.75 & -  &$B(E2;5/2_{3}^{+}\rightarrow 3/2_{2}^{+})$      &0.10   &-\\

&$B(E2;9/2_{4}^{+}\rightarrow 5/2_{3}^{+})$      & 0.76  & -  &$B(E2;7/2_{3}^{+}\rightarrow 5/2_{3}^{+})$      &0.04   &-\\

&$B(E2;11/2_{5}^{+}\rightarrow 7/2_{3}^{+})$      & 1.05   & -  &$B(E2;9/2_{4}^{+}\rightarrow 7/2_{3}^{+})$      &0.02   &-\\

&$B(M1;5/2_{3}^{+}\rightarrow 3/2_{2}^{+})$   & 0.15   & -   &$B(E2;11/2_{5}^{+}\rightarrow 9/2_{4}^{+})$      &0.01   &-\\

&$B(M1;9/2_{4}^{+}\rightarrow 7/2_{3}^{+})$      &0.15   &-  &      &  & \\

\hline
\footnotesize band5 $\rightarrow$ band5
&$B(E2;11/2_{6}^{+}\rightarrow 7/2_{5}^{+})$      & 0.10 & -  &$B(E2;9/2_{5}^{+}\rightarrow 7/2_{5}^{+})$      &0.05   &-\\

&$B(E2;13/2_{7}^{+}\rightarrow 9/2_{5}^{+})$      & 0.25 & -  &$B(E2;11/2_{6}^{+}\rightarrow 9/2_{5}^{+})$      &0.06   &-\\

&$B(E2;15/2_{7}^{+}\rightarrow 11/2_{6}^{+})$      & 0.53   & -   &$B(E2;15/2_{7}^{+}\rightarrow 13/2_{7}^{+})$      &0.03   &-\\

&$B(E2;17/2_{7}^{+}\rightarrow 13/2_{7}^{+})$      & 1.00  & - &$B(E2;17/2_{7}^{+}\rightarrow 15/2_{7}^{+})$      &0.05   &-\\

&$B(M1;9/2_{5}^{+}\rightarrow 7/2_{5}^{+})$    & 0.13 & - &$B(M1;15/2_{7}^{+}\rightarrow 13/2_{7}^{+})$      &0.29   &- \\

&$B(M1;11/2_{6}^{+}\rightarrow 9/2_{5}^{+})$    & 0.21  & - &$B(M1;17/2_{7}^{+}\rightarrow 15/2_{7}^{+})$      &0.30  &- \\

&$B(M1;13/2_{7}^{+}\rightarrow 11/2_{6}^{+})$  & 0.26  & -  &      &  & \\
\hline
\hline
\end{supertabular}
\label{tableGd157-Ne}
\end{center}

\begin{center}
\small
\topcaption{The calculated intraband and interband $E2$ (in units of $e^{2}b^{2}$) and $M1$ (in units of $\mu_{N}^{2}$)
transition rates for low-lying positive-parity bands in $^{157}$Gd.}
\tablefirsthead{\hline  \hline\multicolumn{1}{c}{}&
\multicolumn{1}{c}{} &
   \multicolumn{1}{c}{Th.} &
    \multicolumn{1}{c}{Exp.} &
     \multicolumn{1}{c}{}&
     \multicolumn{1}{c}{Th.} &
      \multicolumn{1}{c}{Exp.} \\\hline}

\tablehead{\multicolumn{3}{l}{continue}\\\hline\hline
\multicolumn{1}{c|}{} &
\multicolumn{1}{c}{} &
   \multicolumn{1}{c}{Th.} &
    \multicolumn{1}{c}{Exp.} &
     \multicolumn{1}{c}{}&
     \multicolumn{1}{c}{Th.} &
      \multicolumn{1}{c}{Exp.} \\\hline}

\begin{supertabular}{p{25mm}|p{42mm}p{11mm}<{\centering}p{15mm}<{\centering}p{42mm}p{11mm}<{\centering}p{15mm}<{\centering}}
\footnotesize band1 $\rightarrow$ band1
&$B(E2;9/2_{1}^{+}\rightarrow 5/2_{1}^{+})$ & 0.49& - &$B(E2;7/2_{1}^{+}\rightarrow 5/2_{1}^{+})$& 1.63&-\\

&$B(E2;11/2_{1}^{+}\rightarrow 7/2_{1}^{+})$ & 0.84  & -  &$B(E2;9/2_{1}^{+}\rightarrow 7/2_{1}^{+})$& 1.33  & -\\

&$B(E2;13/2_{1}^{+}\rightarrow 9/2_{1}^{+})$ &1.08 & - &$B(E2;11/2_{1}^{+}\rightarrow 9/2_{1}^{+})$& 1.01   & -  \\

&$B(E2;15/2_{1}^{+}\rightarrow 11/2_{1}^{+})$  & 1.26 & - &$B(E2;13/2_{1}^{+}\rightarrow 11/2_{1}^{+})$& 0.75   & -  \\

&$B(M1;7/2_{1}^{+}\rightarrow 5/2_{1}^{+})$    &0.07  & - &  $B(E2;15/2_{1}^{+}\rightarrow 13/2_{1}^{+})$&  0.59 & -  \\

&$B(M1;9/2_{1}^{+}\rightarrow 7/2_{1}^{+})$  & 0.11    & - &  $B(M1;13/2_{1}^{+}\rightarrow 11/2_{1}^{+})$& 0.18 & -  \\

&$B(M1;11/2_{1}^{+}\rightarrow 9/2_{1}^{+})$ & 0.13   & - &  $B(M1;15/2_{1}^{+}\rightarrow 13/2_{1}^{+})$& 0.16 & -  \\
\hline
\footnotesize band2 $\rightarrow$ band2
&$B(E2;5/2_{2}^{+}\rightarrow 1/2_{1}^{+})$      & 0.78  & -  &$B(E2;3/2_{1}^{+}\rightarrow 1/2_{1}^{+})$      &0.79  &-\\

&$B(E2;7/2_{2}^{+}\rightarrow 3/2_{1}^{+})$      & 1.01  & -  &$B(E2;5/2_{2}^{+}\rightarrow 3/2_{1}^{+})$      &0.13  &-\\

&$B(E2;9/2_{2}^{+}\rightarrow 5/2_{2}^{+})$      & 1.11   & -  &$B(E2;7/2_{2}^{+}\rightarrow 5/2_{2}^{+})$      &0.08   &-\\

&$B(E2;11/2_{3}^{+}\rightarrow 7/2_{2}^{+})$      & 1.04   & -  &$B(E2;9/2_{2}^{+}\rightarrow 7/2_{2}^{+})$      &0.03   &-\\

&$B(E2;13/2_{2}^{+}\rightarrow 9/2_{2}^{+})$      & 1.30  & -  &$B(E2;11/2_{3}^{+}\rightarrow 9/2_{2}^{+})$      &0.02  &-\\

&$B(M1;9/2_{2}^{+}\rightarrow 7/2_{2}^{+})$      & 0.14  & - &$B(M1;13/2_{2}^{+}\rightarrow 11/2_{3}^{+})$    &0.15   &-\\

\footnotesize band2 $\rightarrow$ band1
&$B(E2;13/2_{2}^{+}\rightarrow 11/2_{1}^{+})$      & 0.01 &  - &    &    & \\
\hline
\footnotesize band3 $\rightarrow$ band3
&$B(E2;5/2_{3}^{+}\rightarrow 1/2_{2}^{+})$      & 0.92   & -  &$B(E2;3/2_{2}^{+}\rightarrow 1/2_{2}^{+})$      &0.92   & -\\

&$B(E2;7/2_{3}^{+}\rightarrow 3/2_{2}^{+})$      & 1.15  & -  &$B(E2;5/2_{3}^{+}\rightarrow 3/2_{2}^{+})$      &0.26   &-\\

&$B(E2;9/2_{3}^{+}\rightarrow 5/2_{3}^{+})$      & 1.34   & -  &$B(E2;7/2_{3}^{+}\rightarrow 5/2_{3}^{+})$      &0.13  &-\\

&$B(E2;11/2_{2}^{+}\rightarrow 7/2_{3}^{+})$      & 1.09   & -  &$B(E2;9/2_{3}^{+}\rightarrow 7/2_{3}^{+})$      &0.08   &-\\

&  $B(M1;9/2_{3}^{+}\rightarrow 7/2_{3}^{+})$    &0.02   &- &$B(E2;11/2_{2}^{+}\rightarrow 9/2_{3}^{+})$      &0.06   &-\\

\footnotesize band3 $\rightarrow$ band2
&$B(E2;7/2_{3}^{+}\rightarrow 3/2_{1}^{+})$   & 0.02  & -  &$B(E2;11/2_{2}^{+}\rightarrow 7/2_{2}^{+})$   & 0.32  & - \\
\hline
\footnotesize band4 $\rightarrow$ band4
&$B(E2;5/2_{5}^{+}\rightarrow 1/2_{3}^{+})$      & 0.20   & -  &$B(E2;3/2_{4}^{+}\rightarrow 1/2_{3}^{+})$      &0.20  & -\\

&$B(E2;7/2_{5}^{+}\rightarrow 3/2_{4}^{+})$      & 0.14  & -  &$B(E2;5/2_{5}^{+}\rightarrow 3/2_{4}^{+})$      &0.04   &-\\

&$B(E2;9/2_{6}^{+}\rightarrow 5/2_{5}^{+})$      & 0.16   & -  &$B(E2;7/2_{5}^{+}\rightarrow 5/2_{5}^{+})$      &0.07 &-\\

&$B(E2;11/2_{6}^{+}\rightarrow 7/2_{5}^{+})$      & 0.05  & -  &$B(E2;9/2_{6}^{+}\rightarrow 7/2_{5}^{+})$      &0.10  &-\\

&$B(M1;3/2_{4}^{+}\rightarrow 1/2_{3}^{+})$    & 0.26  & -  &$B(E2;11/2_{6}^{+}\rightarrow 9/2_{6}^{+})$      &0.32   &-\\

&$B(M1;5/2_{5}^{+}\rightarrow 3/2_{4}^{+})$   & 0.70  & -  &  $B(M1;11/2_{6}^{+}\rightarrow 9/2_{6}^{+})$    &0.70   &-\\

&$B(M1;7/2_{5}^{+}\rightarrow 5/2_{5}^{+})$   & 0.62  &  -  &        &    &  \\

\footnotesize band4 $\rightarrow$ band2
&$B(E2;11/2_{6}^{+}\rightarrow 9/2_{2}^{+})$   & 0.05  & -  & $B(E2;9/2_{6}^{+}\rightarrow 5/2_{2}^{+})$ & 0.04 & - \\
\hline
\end{supertabular}
\label{tableGd157-po}
\end{center}

\begin {table}[h]
\begin {center}
\caption{The probabilities of dominated configurations of selected states in ground state band of $^{159}$Tb.}
\bigskip
\begin {tabular}{cccc}
\hline
\hline
 $J^{\pi}$& $j\otimes R$  &  $A-1$ & $ A+1 $ \\
\hline
$9/2^{+}$  & $2d_{5/2} \otimes 2_{1}^{+}$   &  0.26 &0.07 \\
           & $2d_{5/2} \otimes 6_{1}^{+}$  &  0.40 &0.07 \\
\hline
$11/2^{+}$ & $2d_{5/2} \otimes 4_{1}^{+}$   & 0.34&0.06 \\
           & $2d_{5/2} \otimes 6_{1}^{+}$   &  0.18 &0.06 \\
           & $2d_{5/2} \otimes 8_{1}^{+}$   &  0.15 &0.02 \\
\hline
$13/2^{+}$ & $2d_{5/2} \otimes 4_{1}^{+}$   & 0.26&0.07 \\
           & $2d_{5/2} \otimes 8_{1}^{+}$   &  0.37 &0.06 \\
\hline
$15/2^{+}$ & $2d_{5/2} \otimes 6_{1}^{+}$   & 0.37 &0.07 \\
           & $2d_{5/2} \otimes 8_{1}^{+}$   &  0.14 &0.05 \\
           & $2d_{5/2} \otimes 10_{1}^{+}$   &  0.15&0.02 \\
\hline
$17/2^{+}$ & $2d_{5/2} \otimes 6_{1}^{+}$   & 0.25 &0.07 \\
           & $2d_{5/2} \otimes 10_{1}^{+}$   &  0.36 &0.06 \\
\hline
\end{tabular}
\label{tablewf}
\end{center}
\end{table}

Figure \ref{fig:Tb159EM2} displays the core and single-particle contributions to the intraband $B(E2; J\to J-1)$, $B(E2; J\to J-2)$,
and $B(M1; J\to J-1)$ in the ground state bands  of $^{159}$Tb and $^{157}$Gd. It is found that the $B(E2)$ transitions are dominated by the core component and present monotonically increasing $B(E2; J\to J-2)$ and monotonically decreasing $B(E2; J\to J-1)$ as functions of spin. This is because the core has the majority of the charged particles and they are strongly correlated with the deformation. For the $M1$ transitions, both the core and single particle components contribute to the $B(M1)$ because the $\Delta J=1$ states have a rather large overlap with the dominant configurations ({\it c.f.}, Tab. \ref{tablewf}), and consequently, large reduced matrix elements for both the single particle and core according to Eqs. (\ref{eq:MUsp}-\ref{eq:MVc}, \ref{eq:M1sp}, \ref{eq:M1c}). Moreover, the reduced matrix elements of the two components have the same phase, and therefore leads to an enhancement of the total reduced matrix elements and $B(M1)$.

\section{\label{secIV} Summary and outlook}

In summary, we have developed a microscopic CQC model  for calculating spectroscopic properties of odd-mass nuclei.
The dynamics of our CQC Hamiltonian are determined by
microscopic input energies, quadrupole matrix elements, and pairing gaps corresponding to the collective excitation states of
the even-mass core and spherical single-particle states of the odd nucleon.
These are calculated  from a quadrupole collective Hamiltonian
for collective motion of the core combined with a constrained triaxial relativistic Hartree-Bogoliubov model with a relativistic density functional
PC-PK1 in the particle-hole channel, and a separable pairing force in the particle-particle channel.
In the present version of the model, only the Fermi surface $\lambda$ and coupling strength $\chi$
are specifically adjusted to the experimental data.
The model is tested in a series of illustrative calculations of low-lying spectra
for the axially-deformed odd-proton nucleus $^{159}$Tb and the odd-neutron nucleus $^{157}$Gd.
It can reproduce the excitation energies, kinematic and dynamic moments of inertia, $B(E2)$, as well as the systematic trend of $B(M1)$ very well.
The $\gamma$ phonon excitation bands and the interband $E2$ and $M1$ transition rates are also predicted.
It is also found that  the electric quadrupole transitions are dominated by the core component,
while both the core and single particle components contribute to the magnetic dipole transitions.

In this study, the core quasiparticle coupling is described by a quadrupole interaction with a free parameter $\chi$, and we also find that this parameter is not so easy to to be determined microscopically in the present theoretical framework ({\it c.f.} Appendix \ref{app-A}). This could be modified by using the integral of the intrinsic quasiparticle states (as functions of deformation parameters) calculated from covariant EDF and the wave functions of the core in the collective space as in Ref. \cite{Medjadi86}. The method does not involve any free parameter and can also be easily extended to include the octupole interaction based on our microscopic quadrupole-octupole collective Hamiltonian model \cite{Li13,Li16}. Moreover, the description of $B(M1)$ could be improved by including the polarization effect of the time-reversal breaking using the method recently introduced by Rohozi\'{n}ski \cite{Roho15}.

\appendix
\section{\label{app-A}  Evolution of quasiparticle energies in CQC model}

\begin{figure}[htb]
\includegraphics[scale=0.42]{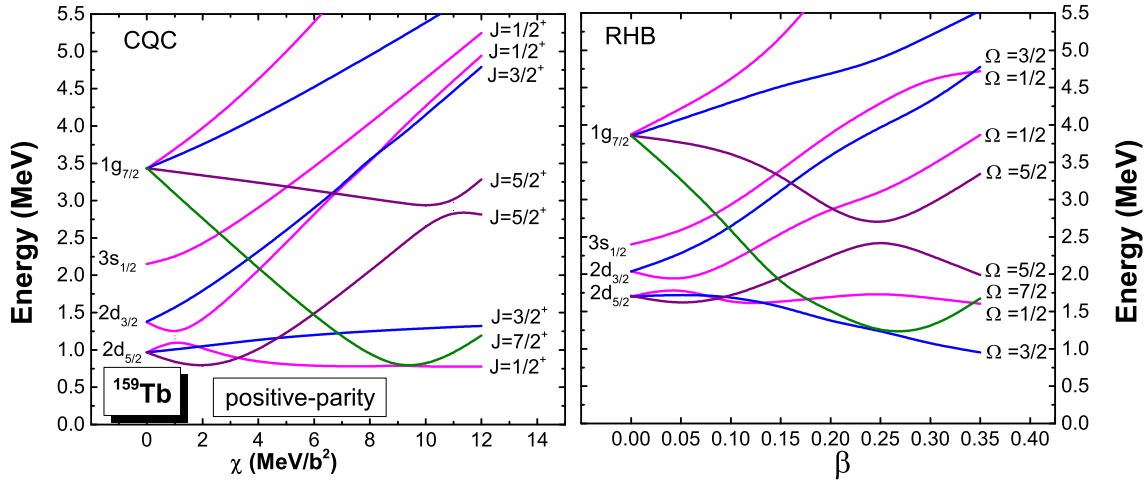}
\caption{(Color online)
(left panel) Quasiparticle energies relevant for the band heads of the low-lying bands as functions of $\chi$ calculated by CQC model with the Fermi surface choosing as the average value of RHB calculation and (right panel) single quasiparticle energies as functions of axially deformed parameter $\beta$ calculated by RHB.}
\label{fig:Tb159single}
\end{figure}

In this part, we take $^{159}$Tb as an example to show the evolution of the quasiparticle energies relevant for the band heads of the low-lying bands (eigen energies of $H_{\rm qp}$ in Eq. (\ref{eq:Ham})) as functions of $\chi$ calculated by CQC model and compare to the single quasiparticle energies as functions of $\beta$ calculated by RHB in Fig. \ref{fig:Tb159single}. The patterns of these two panels are rather similar, while the details are somewhat different, especially for the lowest three levels. This may be because the present CQC model is not self-consistent and does not include higher order multipole interactions. This implies that the two parameters $\lambda$ and $\chi$ are not so easy to be determined microscopically in the present framework of CQC model.

\begin{acknowledgements}
We thank D. Vretenar, T. Nik\v{s}i\'{c}, P. Ring, and J. Li for very helpful discussions.
This work was supported in part by the NSFC under Grants No. 11475140 and No. 11575148,
and the Chinese-Croatian project ``Microscopic Energy Density Functional Theory for Nuclear Fission''.
Work at ORNL is supported by the Office of Nuclear Physics in the U.S. Department of Energy.
\end{acknowledgements}



\end{document}